\documentclass[aps,preprint,groupedaddress,nofootinbib,tightenlines,notitlepage]{revtex4}
\usepackage{amsmath,amssymb}
\usepackage{amsthm}
\usepackage{graphicx}

\usepackage{color}
\usepackage{fancybox}

\newtheorem{Theorem}{Theorem}
\newtheorem{TheoremA}{Theorem}
\newtheorem{Cor}{Corollary}
\newtheorem{CorA}{Corollary}
\newtheorem{Lemma}{Lemma}
\newtheorem{LemmaA}{Lemma}
\newcommand{\Proof}[1][]{\subsection*{#1}}

\def\refa#1{(\ref{#1})}

\def\ra{\rightarrow}

\def\N{\mathbb{N}}
\def\R{\mathbb{R}}
\def\Rb{\overline{\mathbb{R}}}
\def\RTX{\R_+^{1+3}}
\def\RTXa{\R^{1+3}}
\def\RTXb{\Rb_+^{1+3}}
\def\C{\mathcal{C}}

\def\COinf{\mathcal{C}_0^\infty(\R^3)}
\def\D{\mathcal{D}'}
\renewcommand{\L}[1]{{L^{#1}}}
\newcommand{\Lloc}[1][1]{L^{#1}_\text{loc}}
\def\Linf{{L^\infty}}
\def\Linfx#1{{L^\infty_{#1}}}
\def\Linftx#1{{L^\infty_{1,{#1}}}}
\def\LinfX{{L^\infty(\R^3)}}

\def\LinfTXb{{L^\infty(\RTXb)}}
\def\H#1{{\dot{H}^{#1}}}
\def\Hh#1{{\mathcal{H}^{#1}}}
\def\Hd#1{{\dot{\mathcal{H}}^{#1}}}

\def\Dom{\mathcal{D}}

\def\d{\partial}
\newcommand{\dt}[1][]{\frac{d^{#1}}{dt^{#1}}}
\def\lap{\Delta}
\def\la{\lambda}

\def\<{\langle}
\def\>{\rangle}
\def\norm#1{\<#1\>}
\def\nnorm#1{(1+|#1|)} 

\begin{document}

\title{Weighted-$L^\infty$ and pointwise space-time decay estimates\\ 
for wave equations with potentials
and initial data\\ of low regularity}

\author{Nikodem Szpak}
\affiliation{MPI f{\"u}r Gravitationsphysik, Albert-Einstein-Institut\\ 
Am M{\"u}hlenberg 1, 14476 Golm, Germany}

\date{\today}

\begin{abstract} 
We prove weighted-$L^\infty$ and pointwise space-time decay estimates for weak solutions of a class of wave equations with time-independent potentials and subject to initial data, both of low regularity, satisfying given decay bounds at infinity. The rate of their decay depends on the asymptotic behaviour of the potential and of the data. The technique is robust enough to treat also more regular solutions and provides decay estimates for arbitrary derivatives, provided the potential and the data have sufficient regularity, but it is restricted to potentials of bounded strength (such that $-\Delta-|V|$ has no negative eigenvalues).
\end{abstract}


\maketitle

\section{Introduction}

We study a class of wave equations of the form
\begin{equation} \label{wave-eq}
  \Box u + V u := \d^2_t u - \lap u + V(x) u = 0
\end{equation}
where $u:=u(t,x): \Rb_+\times\R^3\ra\R$ and $V(x)$ is a real potential which does not depend on time. We are interested in weak solutions to the initial value problem
\begin{equation} \label{init-data}
  u(0,x)=f(x),\qquad \d_t u(0,x)=g(x)
\end{equation}
with data $(f,g)$ of low regularity satisfying either some weighted-$\Linf$ or pointwise bounds.
We prove a decay estimate in two versions: in stronger, we show a pointwise decay
\begin{equation}
  |u(t,x)| \leq \frac{C}{\nnorm{t+|x|}\nnorm{t-|x|}^{p-1}} \qquad \forall (t,x)\in\Rb_+\times\R^3
\end{equation}
with some $p>2$ provided the potential $V$ and the initial data $f,\nabla f,g$ are continuous and satisfy pointwise bounds
\begin{equation} 
  |V(x)| \leq \frac{V_0}{\nnorm{x}^k},\quad k>2
\end{equation}
with $|V_0|<C_{p,k}^{-1}$ (the value of $C_{p,k}$ will be specified later) and
\begin{equation} 
  |f(x)| \leq \frac{f_0}{\nnorm{x}^{m-1}}, \qquad 
  |\nabla f(x)| \leq \frac{f_1}{\nnorm{x}^m}, \qquad 
  |g(x)| \leq \frac{g_0}{\nnorm{x}^m},\qquad m>3.
\end{equation}
In the weaker version, all pointwise bounds need to hold only \textit{almost everywhere} and so
we prove a weighted-$\Linf$ space-time decay estimate
\begin{equation}
  \|\nnorm{t+|x|}\nnorm{t-|x|}^{p-1} u(t,x)\|_{L^\infty(\Rb_+\times\R^3)} < \infty
\end{equation}
\newpage \noindent
with some $p>2$ provided
\begin{equation} 
  V_0 := \| \nnorm{\cdot}^k V \|_\LinfX < C_{p,k}^{-1} < \infty,\quad k>2
\end{equation}
(the value of $C_{p,k}$ will be specified later) and
\begin{eqnarray} 
  f_0 &:=& \|\nnorm{\cdot}^{m-1} f\|_\LinfX < \infty,\qquad
  f_1 := \|\nnorm{\cdot}^m \nabla f\|_\LinfX < \infty,\qquad \nonumber \\
  g_0 &:=& \|\nnorm{\cdot}^m g\|_\LinfX < \infty,\qquad m>3.
\end{eqnarray}

For both kinds of initial data, $u\in\Lloc(\RTXa)$ represents a weak solution of the wave equation \refa{wave-eq} in the following sense: for any test function $\varphi\in\C^\infty(\RTXa)$
\begin{equation} \label{weak-sol-pot}
  \int dt \int d^3x\; (\Box + V) \varphi(t,x) \; u(t,x) = 
  - \int d^3x\; \d_t \varphi(0,x)\; f(x) + \int d^3x\; \varphi(0,x)\; g(x).
\end{equation}


The initial data $f,\nabla f$ and $g$ have just sufficient spatial decay to have finite energy and we show by functional analytic methods (sec. \ref{Sec-Uniqueness}) that this energy stays finite during the evolution what guarantees uniqueness of the weak solutions.
It follows that the first derivatives $\d_t u, \nabla u\in\L{2} \subset \Lloc$ and hence exist in the weak sense. The second derivatives exist only as distributions, although $\Box u = -V u \in \Linf \subset \Lloc$. Therefore, we work in a subset of the energy space.

There is, in principle, no essential problem with rising the regularity of the solution $u$ provided $f,g$ and $V$ are more regular and their derivatives have proper decay at infinity. Indeed, for $(f,g)\in\C^{n+1}(\R^3)\times\C^n(\R^3)$ and $V\in\C^n(\R^3)$ the solution $u\in\C^n(\RTXb)$ and for $n\geq 2$ it becomes classical. In a separate section (sec. \ref{Sec-Derivatives}) we demonstrate robustness of our technique and show how to obtain estimates for derivatives, provided more assumptions on the initial data and the potential are given. We also easily reproduce the result of Strauss and Tsutaya \cite{Strauss-T} in the case $n=2$.

The technique of proving these estimates relies essentially on the Duhamel integral representation formula for solutions of the free wave equation with a source and non-vanishing initial data. In order for the Duhamel formula to be applicable we need that the functions $f,\nabla f$ and $g$ are Lebesgue integrable over all or almost all spheres in $\R^3$. Therefore, we first choose continuous data and potential, which guarantee the integrability over spheres everywhere. Next, we switch to less regular $\Linf$-spaces where we have integrability over spheres only almost everywhere what, with some additional work regarding the measures of (null) space and space-time subsets, is still sufficient to complete the proof. It seems to be difficult to further generalize the decay theorem by weakening the assumptions and still use the same technique of proof based on the Duhamel representation.

Our motivation for proving these estimates comes mainly from the analysis of nonlinear wave equations, where decay proofs rely on the corresponding results for the linearized equations. We have been also inspired by several nonrigorous approaches to the problem of long-time decay and by numerical observations which consistently show that initial data of compact support (or of enough rapid decay) in presence of a potential $V(x)\sim |x|^{-k}$ ($|x|\gg 1$) evolve to late-time tails of the form $u(t,x)\sim t^{-k}$ for big $t$ and fixed $x$. This has been first explained by Ching \textit{et al.} \cite{ChingComplPRL, ChingTails} who approximated the Green's function or later by Hod \cite{Hod-tails} who used some series expansion, while both arguments were non-rigorous. Strauss and Tsutaya gave in \cite{Strauss-T} a rigorous proof of the decay estimate for classical solutions, i.e. which are twice continuously differentiable, using the Duhamel representation formula. We found the regularity conditions not crucial for controlling the decay of the solution itself and dropped the differentiability conditions for both the data and the potential, thus extending the technique to weak solutions.  By a careful analysis of well-posedness and uniqueness of solutions in the weighted $\Linf$ space, involving proofs of measurability and local integrability combined with functional analytic approach to uniqueness, we were able to prove the same decay estimates (without or with control of the derivatives) and energy conservation for rough solutions.
This setting covers various cases of cut-off potentials or cut-off initial data often discussed in the context of specific approximations or in numerical calculations which were not covered by the existing literature.

\subsection*{Strategy of proof}

We introduce the following notation for solutions of the wave equations. Let $I_V$ be a linear map from the space of initial data to the space of solutions of the wave equation \refa{wave-eq}-\refa{init-data}, so that $u=I_V(f,g)$. For wave equations with a source term and null initial data
\begin{equation} 
  \Box u + Vu= F,\qquad u(0,x)=0,\qquad  \d_t u(0,x)=0,
\end{equation}
let's denote the solutions by $u=L_V(F)$, where $L_V$ is a linear map from the space of source functions to the space of solutions to the above problem. Due to linearity, the solution $u$ of a wave equation with source $F$ and non-vanishing initial data $f,g$ is a sum of these two contributions
\begin{equation}
  u=L_V(F)+I_V(f,g).
\end{equation}
The method we use for proving decay of solutions in presence of a potential is to treat the potential term as a source term. If we put the term $Vu$ in \refa{wave-eq} on the r.h.s
\begin{equation} 
  \Box u = -V u,
\end{equation}
we get a pseudo-free wave equation with a source term $F\equiv -Vu$. Hence, the solution must satisfy
\begin{equation} 
  u = I_V(f,g) = L_0(-Vu)+I_0(f,g),
\end{equation}
where the difficulty is that $u$ appears on both sides. By introducing a weighted space-time norm $\|\chi\cdot\|$ (either standard or essential supremum), with some weight $\chi(t,x)$ (to be specified later), we are able to bound the terms containing $u$ on both sides of the previous equation in such a way that 
\begin{equation}
  \|\chi u\| \leq \delta(V) \|\chi u\| + C(f,g).
\end{equation}
Then if $\delta(V)<1$ we arrive at
\begin{equation}
  \|\chi u\|<\frac{C(f,g)}{1-\delta(V)} =: \widetilde{C}(f,g,V).
\end{equation}
In the case of pointwise bounds the $\|\chi\cdot\|$ norm is a standard weighted supremum norm, which implies a pointwise decay estimate
\begin{equation}
  |u(t,x)| \leq \frac{\widetilde{C}(f,g,V)}{|\chi(t,x)|}.
\end{equation}
In the case of $\Linf$ norms, this inequality holds almost everywhere. 

The condition $\delta(V)<1$ presents a bound on the strength of the potential $|V|$ and is a necessary assumption restricting the theorem. Although such a bound is not expected for non-negative potentials\footnote{There still does not exist an analogous decay theorem, there is only strong numerical evidence. In \cite{Georg-H-K} some weaker decay is proved for arbitrarily strong positive potentials.} the reason for the restriction of $|V|$ is that the technique of proof does not distinguish between negative and positive potentials. Therefore, everything what can be proved, must hold for negative potentials of the given strength $|V|$. Yet, negative potentials may have bound states (i.e. eigenfunctions of $-\lap+V$) which may grow exponentially in time, thus destroying the decay. Hence, the condition $|V(x)|\leq V_0/\norm{x}^k$ with $k>2$ and $V_0$ small enough must in fact avoid bound states. Indeed, this can be observed here, at least for spherically symmetric potentials where several bounds on the number of bound states are well-known \cite[Th. XIII.9]{RS-IV}, e.g. the Bargmann's bound
\begin{equation}
  N\leq \int_0^\infty r |V(r)| dr \leq \frac{V_0}{(k-1)(k-2)}
\end{equation}
or the Calogero's bound
\begin{equation}
  N\leq \frac{2}{\pi}\int_0^\infty \sqrt{|V(r)|} dr \leq \frac{2}{\pi}\frac{\sqrt{V_0}}{(k-2)}.
\end{equation}
We get $N=0$ if $N<1$, i.e. if $V_0<(k-1)(k-2)$ or $V_0<\frac{\pi^2}{16} (k-2)^2$. The bound, which appears in our proofs $V_0<C_{p,k}^{-1}$, although far from optimal, guarantees these conditions. 

We make an effort to estimate the numerical value of the constant $C_{p,k}$ since in some applications the potential appears with a given value of $V_0$ which is not (arbitrarily) small, e.g. waves propagating in the Schwarzschild geometry give rise to an effective potential.


\begin{acknowledgments}
I want to express my deep gratitude for Roger Bieli who contributed a lot to the proofs regarding measurability and integrability of weak solutions (lemmas \ref{Lem-L-free_wave_eq} and \ref{Lem-L-Georgiev}) and read carefully the manuscript. I also want to thank to Andrzej Herdegen and Horst Beyer for valuable comments on the evolution in Hilbert spaces and to Michael Beals  for an interesting discussion.
\end{acknowledgments}

\section{Main results}
\subsection*{Definitions}

We denote by $\R_+:=(0,\infty)$ and its closure by $\Rb_+:=[0,\infty)$. Analogously, we denote by $\RTX:=\R_+\times\R^3$ and by $\RTXb:=\Rb_+\times\R^3$. 

We denote by $S(x,t)$ a sphere in $\R^3$ with the center $x$ and radius $t$, i.e. $S(x,t):=\{y\in\R^3: |y-x|=t\}$, and by $K(x,t)$ a (past-)cone in $\RTX$ with the center at point $(t,x)$, i.e. $K(x,t):=\{(s,y)\in\RTX: |y-x|=t-s\}$. The corresponding Lebesgue measures on $S(x,t)$ and $K(x,t)$ we denote by $d\sigma(y)$ with $y\in S(x,t)$ and $d\kappa(s,y)$ with $(s,y)\in \Rb\times \R^3 \cap K(x,t)$, respectively.

We introduce a frequently used short hand
\begin{equation}
  \<x\>:=1+|x|.
\end{equation}
We remind that the space $\Linf(\R^n)$ consists of Lebesgue measurable functions on $\R^n$ having finite $\Linf$ norm which is defined as
\begin{equation}
  \|h\|_\Linf := \mathop{\text{ess-sup}}_{x\in\R^n} |h(x)| 
  \equiv \inf\{M: |h(x)|\leq M \text{ a.e. on } \R^n\}.
\end{equation}
Here a.e. on $\R^n$ means almost everywhere on $\R^n$ in the sense that it does not hold at most on a set of measure zero in $\R^n$ with the standard Lebesgue measure on $\R^n$. For bounded functions $h(x)$ and $H(t,x)$ we will use the notation
\begin{equation}
  \|h\|_\infty := \sup_{x\in\R^3} |h(x)|\qquad\text{ and }\qquad
  \|H\|_\infty := \sup_{(t,x)\in\RTXb} |H(t,x)|
\end{equation}
identical in both cases if it does not lead to confusion (otherwise we will use the symbol $\sup$ explicitly). We define weighted-$\Linf$ spaces for measurable functions defined on space ($\R^3$) or space-time ($\RTXb$)
\begin{equation}
  h\in\Linfx{\mu}\quad\Leftrightarrow\quad \|h\|_\Linfx{\mu} := 
  \|\<\cdot\>^\mu h\|_{L^\infty(\R^3)} < \infty,
\end{equation}
\begin{equation}
  H\in L^\infty_{r,p}\quad\Leftrightarrow\quad \|H\|_{L^\infty_{r,p}} := 
  \|\<t+|x|\>^r\<t-|x|\>^{p-r} H(t,x)\|_{L^\infty_t(\Rb_+) L^\infty_x(\R^3)} < \infty.
\end{equation}
The content of the last definition is that $H(t,x)$ decays like $1/t^r$ along the lightcone and like $1/t^p$ or $1/|x|^p$ at timelike or spatial infinity, respectively. We will mainly consider spaces $\Linftx{p}$.
We define some constants used throughout the article
\begin{equation}
  c_p:=1/2(p-2),
\end{equation}
\begin{equation}
  C^{(1)}_p := \max \left(\frac{9}{2(p-2)}, 4\right),\qquad
  C^{(2)}_p := \max \left(\frac{3}{p-1}, 5\right),
\end{equation}
\begin{equation}
  C_m:= \max(C^{(1)}_m,C^{(2)}_m) = \max \left(\frac{9}{2(m-2)}, 5\right),
\end{equation}
\begin{equation}
  C_{p,q}:=\frac{3}{2} \frac{6^{q-1}}{(q-2)} \max(2/(p-1),3).
\end{equation}

\subsection*{Theorems}

First, existence and uniqueness of weak solutions is proved together with conservation of the energy. The theorem provides also $\L{2}$--bounds on $\nabla u(t)$ and $\d_t u(t)$ which are not contained in the following decay theorems (because the derivatives there are considered only to exist in the distributional sense) and implies that in fact the first derivatives exist in the weak sense. 
\begin{Theorem}[Existence and uniqueness] \label{Th-Energy}
With the assumptions of either Theorem \ref{Th-decay} or \ref{Th-Ldecay} the (weak) solution $u$ of the wave equation \refa{wave-eq}-\refa{init-data} is unique, belongs to $\C^0(\Rb_+,H^1)\cap C^1(\Rb_+,\L{2})$ and gives rise to a conserved energy
\begin{equation}
\begin{split}
  E[u(t)] &:= \<\nabla u(t), \nabla u(t)\> + \<u(t), V u(t)\> + \<\d_t u(t), \d_t u(t)\>\\
  = E[u(0)] &= \<\nabla f, \nabla f\> + \<f, V f\> + \<g, g\> .
\end{split}
\end{equation}
Moreover, $\nabla u(t), \d_t u(t) \in \L{2}$ for all $t\in\Rb_+$ what implies that the first derivatives exist in a weak sense, i.e. $\nabla u(t), \d_t u(t) \in \Lloc$ for all $t\in\Rb_+$.
\end{Theorem}
The case when additionally $V\geq 0$ is somewhat simpler to treat and Thoe \cite{Thoe-WavPot} has given a complete spectral theory including explicit representation of the evolution operator in action on the generalized eigenfunctions. It provides conservation of energy, but it does not seem to be of a big advantage when proving decay estimates of the type we are interested in.

It is important to note that for continuous potentials there is no need of using such sophisticated tools like functional analysis for proving uniqueness of classical solutions. In this case the standard energy inequality is sufficient. Indeed, note first that $V\in\C^2$ is enough to guarantee the regularity of classical solutions $u\in\C^2$ and by differentiation of the free energy $E_0(t):=\|\nabla u\|_\L{2}^2+\|\dot{u}\|_\L{2}^2$ with respect to time one obtains easily 
\begin{equation}
  \frac{dE_0(t)}{dt} \leq 2 V_0 E_0(t),
\end{equation}
what implies $E_0(t)\leq E_0(0) \exp(2 V_0 t)$. This, together with the bound $\|u(t)\|_\L{2}\leq \|u(0)\|_\L{2} + \int_0^t \|\dot{u}(t')\|_\L{2}\; dt'$, gives bounds on $(u(t),\dot{u}(t))\in H^1\times\L{2}$ in terms of the initial data. By the density argument these bounds remain true in case $V,u\in \C^0$ (with weak derivatives and finite energy initial data), what guarantees uniqueness of solutions in $\C^0(\Rb_+,H^1)\cap \C^1(\Rb_+,\L{2})$. Unfortunately, this technique cannot be extended to the case of weighted $\Linf$ spaces, because $\C^n$ are not dense in $\Linf$.

We state the results regarding the continuous pointwise and weighted-$\Linf$ estimates separately. The first theorem contains a pointwise estimate for weak solutions which are only continuous and presents a generalization of the theorem of Strauss \textit{et al.} \cite{Strauss-T} for classical solutions (twice continuously differentiable).
\renewcommand{\theTheoremA}{2a}
\begin{TheoremA}[Decay in $\C^0$] \label{Th-decay}
Let $(f,g)\in \C^1(\R^3)\times\C^0(\R^3)$ and $V\in\C^0(\R^3)$. If for some $k>2$ 
\begin{equation} \label{V-bound}
  |V(x)| \leq \frac{V_0}{\norm{x}^k}\qquad\forall x\in\R^3 
  \qquad\text{with}\qquad V_0< C_{p,k}^{-1}<\infty 
\end{equation}
and for some $m>3$
\begin{equation} \label{f-g-bound}
  |f(x)| \leq \frac{f_0}{\norm{x}^{m-1}},\qquad 
  |\nabla f(x)| \leq \frac{f_1}{\norm{x}^m},\qquad 
  |g(x)| \leq \frac{g_0}{\norm{x}^m},\qquad\forall x\in\R^3
\end{equation}
then the wave equation \refa{wave-eq} with the initial data \refa{init-data} has a unique weak solution $u\in\C^0(\RTXb)$ which satisfies
\begin{equation}
  |u(t,x)| \leq \frac{C}{\norm{t+|x|}\norm{t-|x|}^{p-1}},\qquad \forall (t,x)\in\RTXb
\end{equation}
with $p:=\min(k,m-1)$ and 
\begin{equation}
  C:=\frac{C_m(f_0+f_1+g_0)}{1-C_{p,k} V_0}.
\end{equation}
\end{TheoremA}

The condition $|f(x)|\leq f_0/\norm{x}^{m-1}$ can be replaced by a weaker $|f(x)|\ra 0$ as $|x|\ra \infty$. The pointwise estimate follows then from integration of $|\nabla f(x)|\leq f_1/\norm{x}^{m}$ from infinity to $x$ along the radial direction. Therefore, it holds $f_0\leq f_1/(m\, 2^m)$.

The second theorem generalizes the estimate further to weak solutions which belong only to a weighted-$\Linf$ space. This seems to be the widest space compatible with the assumptions of spatial decay of the initial data and the potential.
\renewcommand{\theTheoremA}{2b}
\begin{TheoremA}[Decay in $\Linftx{p}$] \label{Th-Ldecay}
If for some $k>2$ 
\begin{equation} \label{V-Lbound}
  V_0:=\|V\|_\Linfx{k} < C_{p,k}^{-1} < \infty,
\end{equation}
and for some $m>3$
\begin{equation} \label{f-g-Lbound}
  f_0:=\|f\|_\Linfx{m-1} < \infty \qquad
  f_1:=\|\nabla f\|_\Linfx{m} < \infty \qquad
  g_0:=\|g\|_\Linfx{m} < \infty
\end{equation}
then the wave equation \refa{wave-eq} with the initial data \refa{init-data} has a unique weak solution which satisfies
\begin{equation}
  \|u\|_\Linftx{p} \leq C,
\end{equation}
with $p:=\min(k,m-1)$ and 
\begin{equation}
  C:=\frac{C_m(f_0+f_1+g_0)}{1-C_{p,k} V_0}.
\end{equation}
\end{TheoremA}
\addtocounter{Theorem}{1}
\noindent Note that the estimates of Theorem \ref{Th-decay} or \ref{Th-Ldecay}
\begin{equation}
  |u(t,x)| \leq \frac{C}{\norm{t+|x|}\norm{t-|x|}^{p-1}} \leq 
  \frac{C(t)}{\norm{x}^{p}}, \qquad \text{a.e. for }x\in\R^3, \quad \forall t\in\Rb_+
\end{equation}
for $p>2$ and some $C(t)<\infty$ for all $t\in\Rb_+$ imply $u(t)\in\L{2}$ for all $t\geq 0$.

Then, in a straightforward way we can prove the following, useful for further applications,
\begin{Cor}[Decay in presence of a source] \label{Cor-wave-eq-V-F}
Let the initial data satisfy 
\begin{equation}
  f_0:=\|f\|_\Linfx{m-1} < \infty,\qquad
  f_1:=\|\nabla f\|_\Linfx{m} < \infty,\qquad
  g_0:=\|g\|_\Linfx{m} < \infty
\end{equation}
with $m>3$, the potential satisfy
\begin{equation}
  V_0:=\|V\|_\Linfx{k} < C_{p,k}^{-1} < \infty
\end{equation}
with $k>2$ and the source $F\in\Linftx{r}$ satisfy for some $q>2$ and $1<r\leq q$
\begin{equation}
  F_0:=\|\norm{x}^q F\|_\Linftx{r} < \infty.
\end{equation}
Then there exists a weak solution $u(t,x)$ of the wave equation with potential and source terms
\begin{equation} \label{wave-eq-V-F}
  \Box u + V u = F,\qquad u(0,x)=f(x),\qquad \d_t u(0,x)=g(x)
\end{equation}
and initial data $u(0,x)=f(x)$, $\d_t u(0,x)=g(x)$ which satisfies
\begin{equation}
  \|u\|_\Linftx{p}\leq \frac{C_m(f_0+f_1+g_0)+C_{r,q} F_0}{1-C_{p,k} V_0}.
\end{equation}
for $p:=\min(k,m-1,r)$.

If additionally $(f,g)\in\C^1(\R^3)\cap\C^0(\R^3)$, $V\in\C^0(\R^3)$ and $F\in\C^0(\RTXb)$ then $u\in\C^0(\RTXb)$ and the same estimates hold pointwise\footnote{i.e. everywhere, instead of almost everywhere.}.
\end{Cor}

\begin{Cor}[Decay of derivatives] \label{Cor-decay-derivs}
The weak solution of the wave equation \refa{wave-eq} with the initial data \refa{init-data} satisfies
\begin{equation}
  u, |\nabla u|, |\nabla^2 u|, \d_t u, \d_t^2 u \in \Linftx{p}  
\end{equation}
with $p:=\min(k,m-1)$ provided
\begin{equation}
  f, |\nabla f|, |\nabla^2 f|, |\nabla^3 f|, g, |\nabla g|, |\nabla^2 g| \in \Linfx{m},\qquad m>3
\end{equation}
\begin{equation}
  V, |\nabla V|, |\nabla ^2 V| \in \Linfx{k},\quad k>2 \quad \text{and} \quad
  \|V\|_\Linfx{k} < C_{p,k}^{-1}< \infty.
\end{equation}
If additionally $(f,g)\in\C^3(\R^3)\cap\C^2(\R^3)$ and $V\in\C^2(\R^3)$ then $u\in\C^2(\RTXb)$ and the estimates hold pointwise\footnotemark[\value{footnote}].
\end{Cor}
\noindent Here, we have introduced notation $|\nabla^n h|:=\sum_{a_1,...,a_n=1}^3 |\d_{a_1} ... \d_{a_n} h|$. This is the same result (in the $u\in\C^2$ case) which was obtained by Strauss and Tsutaya in \cite{Strauss-T}.

\section{Basic estimates} 

The first lemma estimates an integral appearing in the Duhamel representation formula containing initial data. It was proved in somewhat different form by many authors \cite{Strauss-T, Asakura, John-blowup}, but here we prove only a simplified version, which is needed for our goals. Actually, this estimate can be still improved by one power of $\norm{t-|x|}$, what better applies to long range initial data, which are not of our main interest.
\begin{Lemma} \label{Lem-int}
For $t>0$, $p>2$ and
\begin{equation} \label{lem1-I1}
  I:= \frac{1}{4\pi}\int_{S(x,t)} \frac{d\sigma(y)}{\norm{y}^p}
\end{equation}
it holds
\begin{equation} \label{lem1-est}
  I\leq c_p\, \frac{t}{|x|\norm{t-|x|}^{p-2}}.
\end{equation}
Moreover, for $p>2$
\begin{equation} \label{est-int1}
  \frac{1}{4\pi t} \int_{S(x,t)} \frac{d\sigma(y)}{\norm{y}^p}
  \leq \frac{C^{(1)}_p}{\norm{t+|x|}\norm{t-|x|}^{p-2}}
\end{equation}
and for $p>3$
\begin{equation} \label{est-int2}
  \frac{1}{4\pi t^2} \int_{S(x,t)} \frac{d\sigma(y)}{\norm{y}^{p-1}}
  \leq \frac{C^{(2)}_p}{\norm{t+|x|}\norm{t-|x|}^{p-2}}.
\end{equation}
\end{Lemma}

The next lemma estimates an integral appearing in the representation formula containing a source term. We follow the proof of Georgiev \textit{et al.} \cite{Georg-H-K}. Similar calculation has been done by Strauss \textit{et al.} in \cite{Strauss-T} and earlier by John \cite{John-blowup}.
\begin{Lemma} \label{Lem-int-cone}
  For $t>0$, $x\in\R^3$ and
\begin{equation}
  I:=\frac{1}{4\pi}\int_{K(x,t)} \frac{d\kappa(s,y)}{(t-s) \norm{y}^q \norm{s+|y|} \norm{s-|y|}^{p-1}}
\end{equation}
for $q>2$ and $q\geq p>1$ it holds
\begin{equation}
  |I|\leq \frac{C_{p,q}}{\norm{t+|x|} \norm{t-|x|}^{p-1}}.
\end{equation}
\end{Lemma}

Next, we prove the general representation formula for distributional and weak solutions.
\begin{Lemma} \label{Lem-weak-sol}
  For any distributions  $f,g\in\D(\R^3)$ and $F\in\D(\RTXa)$ supported on $\RTXb$ the formula 
\begin{equation} \label{lem-Duhamel-conv}
\begin{split}
  v(t,x)&=\d_t \int_{\R^3} \frac{\delta(t-|x-y|)}{4\pi t} f(y) d^3y
  + \int_{\R^3} \frac{\delta(t-|x-y|)}{4\pi t} g(y) d^3y \\
  &+ \int_{\Rb_+} ds \int_{\R^3} d^3y\; \frac{\delta(t-s-|x-y|)}{4\pi (t-s)} F(s,y)
\end{split}
\end{equation}
defines a distribution $v\in\D(\RTXa)$ supported on $\RTXb$ which satisfies the wave equation 
\begin{equation} \label{wave-eq-dist}
  \Box v = F,\qquad v(0,x)=f(x),\qquad \d_t v(0,x)=g(x)
\end{equation}  
in the weak sense, i.e. for any test function $\varphi\in\C_0^\infty(\R^{1+3})$
\begin{equation} \label{weak-sol-lem}
\begin{split}
  \int dt \int d^3x\; \Box \varphi(t,x) \; v(t,x) =& 
  - \int d^3x\; \d_t \varphi(0,x)\; f(x) + \int d^3x\; \varphi(0,x)\; g(x)\\
  &+ \int_{\R} dt \int d^3x\;  \varphi(t,x)\; F(t,x).
 \end{split}
\end{equation}
If $f,g,F$ are ordinary functions (i.e. regular distributions) or in $\Lloc$ and $v\in\Lloc$ is defined almost everywhere in $\RTXb$ by the Duhamel's formula 
\begin{equation} \label{lem-Duhamel}
\begin{split}
  v(t,x)&=\frac{1}{4\pi}\int_{S(x,t)} \frac{g(y)}{t}\; d\sigma(y)
  +\frac{1}{4\pi}\int_{S(x,t)} \frac{(y-x)\cdot\nabla f(y)+f(y)}{t^2}\; d\sigma(y)\\
  &+\frac{1}{4\pi} \int_{K(x,t)} \frac{F(s,y)}{t-s}\; d\kappa(s,y).
\end{split}
\end{equation}
then $v$ is a weak solution, in the sense specified above, of the wave equation \refa{wave-eq-dist}.
\end{Lemma}

The following lemma, an estimate for solutions to the free wave equation with prescribed initial data, has been proved by Strauss \textit{et al.} \cite{Strauss-T} for classical solutions, i.e. for $(f,g)\in\C^3(\R^3)\times \C^2(\R^3)$ leading to $u\in\C^2(\RTXb)$. We weaken the assumptions to $(f,g)\in\C^1(\R^3)\times \C^0(\R^3)$ and obtain $u\in\C^0(\RTXb)$.
\addtocounter{Lemma}{1}
\renewcommand{\theLemmaA}{\arabic{Lemma}a}
\begin{LemmaA} \label{Lem-free_wave_eq}
Let the data $(f,g)\in\C^1(\R^3)\times\C^0(\R^3)$ and satisfy 
\begin{equation} 
  |f(x)| \leq \frac{f_0}{\norm{x}^{m-1}},\qquad
  |\nabla f(x)| \leq \frac{f_1}{\norm{x}^m},\qquad
  |g(x)| \leq \frac{g_0}{\norm{x}^m},\qquad \forall x\in\R^3
\end{equation}
for some $m>3$. Then there exists a unique weak solution $v(t,x)=I_0(f,g)$ of the free wave equation
\begin{equation} \label{lem-wave-eq}
  \Box v = 0,\qquad v(0,x)=f(x),\qquad \d_t v(0,x)=g(x).
\end{equation}  
Moreover, it is continuous in $(t,x)\in\RTXb$ and satisfies
\begin{equation}
  |v(t,x)| \leq \frac{C(f,g)}{\norm{t+|x|}\norm{t-|x|}^{m-2}}\qquad \forall (t,x)\in\RTXb,
\end{equation}
where $C(f,g):=C_m\cdot (g_0+f_1+f_0)$.
\end{LemmaA}

As next, we further weaken the assumptions to $\nabla f,g\in\Linfx{m}$ and $f\in\Linfx{m-1}$, i.e. to weighted $\LinfX$ spaces.
\renewcommand{\theLemmaA}{\arabic{Lemma}b}
\begin{LemmaA} \label{Lem-L-free_wave_eq}
Let the data $(f,g)\in\Linfx{m-1}\times\Linfx{m}$ with $m>3$ satisfy 
\begin{equation}
  f_0:=\|f\|_\Linfx{m-1} < \infty,\qquad
  f_1:=\|\nabla f\|_\Linfx{m} < \infty,\qquad
  g_0:=\|g\|_\Linfx{m} < \infty.
\end{equation}
Then there exists a unique weak solution $v(t,x)=I_0(f,g)$ of the free wave equation
\begin{equation} \label{lem-wave-eq-L}
  \Box v = 0,\qquad v(0,x)=f(x),\qquad \d_t v(0,x)=g(x)
\end{equation}  
which satisfies
\begin{equation}
  \|v\|_\Linftx{m-1} \leq C(f,g) := C_m\cdot (g_0+f_1+f_0).
\end{equation}
\end{LemmaA}

The following lemma, an estimate for solutions to the wave equation with source, has been proved by Strauss \textit{et al.} \cite{Strauss-T} and Asakura \cite{Asakura} for classical solutions, i.e. for $F\in\C^2(\RTXb)$ leading to $u\in\C^2(\RTXb)$. We weaken the assumption to $F\in\C^0(\RTXb)$ and obtain $u\in\C^0(\RTXb)$.
\addtocounter{Lemma}{1}
\renewcommand{\theLemmaA}{\arabic{Lemma}a}
\begin{LemmaA} \label{Lem-Georgiev}
Let the source $F\in\C^0(\RTXb)$ and satisfy for some $q>2$ and $1<p\leq q$
\begin{equation}
  |F(t,x)| \leq \frac{F_0}{\norm{t+|x|} \norm{t-|x|}^{p-1} \norm{x}^q}\qquad \forall (t,x)\in\RTXb.
\end{equation}
Then there exists a weak solution $v(t,x)=L_0(F)$ of the free wave equation with source
\begin{equation} \label{wave-eq-F}
  \Box v = F,
\end{equation}
and null initial data $v(0,x)=0$, $\d_t v(0,x)=0$. Moreover, it is continuous in $(t,x)\in\RTXb$ and satisfies
\begin{equation}
  |v(t,x)| \leq \frac{C_{p,q} F_0}{\norm{t+|x|}\norm{t-|x|}^{p-1}}\qquad \forall (t,x)\in\RTXb.
\end{equation}
\end{LemmaA}

As next, we weaken the assumption to $F\in\norm{x}^{-q}\Linftx{p}$, i.e. a weighted $\LinfTXb$ space.
\renewcommand{\theLemmaA}{\arabic{Lemma}b}
\begin{LemmaA} \label{Lem-L-Georgiev}
Let the source $F$ satisfy for some $q>2$ and $1<p\leq q$
\begin{equation}
  F_0:=\|\norm{x}^q F\|_\Linftx{p} < \infty.
\end{equation}
Then there exists a weak solution $v(t,x)=L_0(F)$ of the free wave equation with source
\begin{equation} \label{wave-eq-F-L}
  \Box v = F,
\end{equation}
and null initial data $v(0,x)=0$, $\d_t v(0,x)=0$. Moreover, it satisfies
\begin{equation}
  \|v\|_\Linftx{p} \leq C_{p,q} F_0.
\end{equation}
\end{LemmaA}


\section{Existence and uniqueness in the energy space} \label{Sec-Uniqueness}

For the free wave equation with $V\equiv 0$ it is well known that the homogeneous Sobolev spaces $\H{s}(\R^3)\times\H{s-1}(\R^3)$ have the property that the evolution operator $W(t)$ generating the solution $(u(t),\d_t u(t)) = W(t) (f,g)$ of the free wave equation is unitary and the norm defines a naturally conserved energy
\begin{equation}
  E_s[u(t)] := \|u\|^2_\H{s} + \|\d_t u\|^2_\H{s-1}
\end{equation}
with the case $s=1$ of our special interest
\begin{equation}
  E_0[u(t)] := \int\left(|\nabla u|^2 + |\d_t u|^2 \right) d^3x.
\end{equation}
This is no more true in presence of the potential $V$. Yet, if the potential satisfies some weakness conditions we show that there exist perturbed homogeneous Sobolev spaces $\Hd{s}(\R^3)\times\Hd{s-1}(\R^3)$ in which the evolution is unitary. 

We start from the observation that initial data $(f,g)$ satisfying the decay bounds \refa{f-g-bound} or \refa{f-g-Lbound} of Theorem \ref{Th-decay} or \ref{Th-Ldecay}, respectively, 
\begin{equation} \label{f-g-bound-App}
  |f(x)| \leq \frac{f_0}{\norm{x}^{m-1}},\qquad 
  |\nabla f(x)| \leq \frac{f_1}{\norm{x}^m},\qquad 
  |g(x)| \leq \frac{g_0}{\norm{x}^m},\qquad m>3,
\end{equation}
holding either everywhere or almost everywhere in $\R^3$, belong to the following spaces: $X_0:=\H{1}\times\L{2}$ (the free energy space), $X_1:=H^1\times\L{2}$ and $X_E:=\Hd{s}\times \Hd{s-1}$ (defined below) with $s=1$, which we will call the (true) energy space for the problem at hand. In the following we will prove that if the potential satisfies the bound \refa{V-bound} or \refa{V-Lbound} of Theorem \ref{Th-decay} or \ref{Th-Ldecay}, respectively, 
\begin{equation} \label{V-bound-App}
  |V(x)| \leq \frac{V_0}{\norm{x}^k}\qquad k>2,
\end{equation}
again everywhere or almost everywhere in $\R^3$, then there always exists a unique solution $(u,\dot{u})\in\C(\R_+,X_0), \C(\R_+,X_1)$ or $\C(\R_+,X_E)$ of the wave equation \refa{wave-eq} for any initial data $(f,g)\in X_0, X_1$ or $X_E$, respectively, but only in the true energy space $X_E$ the evolution is unitary and the corresponding energy conserved when $V\neq  0$.

In dimension $n=3$ the bounds \refa{f-g-bound-App} and \refa{V-bound-App} imply
\begin{equation} \label{f-g-L2}
    f, \nabla f, g\in \L{2} \qquad \text{and}\qquad V\in \L{2}.
\end{equation}
Moreover, from \refa{V-bound-App} with $k>2$ follows
\begin{equation}
  |V(x)| \leq \frac{V_0}{(1+|x|)^2} \leq \frac{V_0}{|x|^2}
\end{equation}
and with $\<\cdot, \cdot\>$, being the standard scalar product in $\L{2}$, we have for any $h\in H^{1}$
\begin{equation} \label{fVf-Hardy}
  |\<h,Vh\>| = \left| \int V(x) |h(x)|^2 d^3x \right| \leq V_0 \int \frac{|h(x)|^2}{|x|^2} d^3x 
  \leq 4 V_0 \int |\nabla h|^2 d^3x = 4 V_0 \|\nabla h\|_\L{2}^2
\end{equation}
by Hardy's inequality for $n=3$. For any $h\in\L{2}$ we have
\begin{equation} \label{fVf-L2}
  |\<h,Vh\>| = \left| \int V(x) |h(x)|^2 d^3x \right| \leq  \| V \|_\Linf \| h \|_\L{2}^2
  \leq V_0\, \| h \|_\L{2}^2.
\end{equation}
For the classical solutions $u\in\C^2(\RTXb)$ we know that the energy, defined by
\begin{equation}
  E[u(t)] := \int\left(|\nabla u|^2 + |\d_t u|^2 + V |u|^2\right) d^3x,
\end{equation}
is conserved in evolution, what can be proved by differentiation with respect to time. This technique does not work for weak solutions, for which the derivatives exist only as distributions. For our rough data $(f,g)$ we have from \refa{fVf-L2}
\begin{equation} \label{E-finite}
  \left|E[u(0)]\right| = \left|\int\left(|\nabla f|^2 + |g|^2 + V |f|^2\right) d^3x\right|
  \leq \|\nabla f\|_\L{2}^2 + \|g\|_\L{2}^2 + V_0\, \|f\|_\L{2}^2 < \infty.
\end{equation}
This energy is positive definite for potentials which are weak enough\footnote{It is also positive for all $V(x)\geq 0$, but here we are interested only in bounds of the form \refa{V-bound-App}} and satisfy $V_0<\frac{1}{4}$, what follows from \refa{fVf-Hardy} and
\begin{equation} \label{Df-fVf-Df}
  \|\nabla f\|^2_\L{2} + \<f,Vf\> \geq \|\nabla f\|^2_\L{2} -  4 V_0 \int |\nabla f|^2 d^3x 
  = (1-4 V_0) \|\nabla f\|^2_\L{2} \geq 0.
\end{equation}
In this section we will show that there is a unique evolution in the space of solutions with finite energy.

First, consider $A:=-\lap+V$ defined on $\COinf$. For $V\in\Linf\cap\L{2}$, what is guaranteed by \refa{V-bound-App}, the operator $A$ is essentially self-adjoint on $\C^\infty_0(\R^3)$ and has a unique self-adjoint extension (which we denote for simplicity again by $A$) to $\Dom(A)=\Dom(-\lap)=\{f\in\L{2}: \lap f\in\L{2}\}=H^2$ (see e.g. \cite[Th. X.15]{RS-II}).

For $f\in H^2$ we can partially integrate and find as in \refa{Df-fVf-Df} that
\begin{equation}
  \<f,Af\> = \<f,-\lap f\> + \<f,Vf\> = \|\nabla f\|^2_\L{2} + \<f,Vf\> 
  \geq (1-4 V_0) \|\nabla f\|^2_\L{2} \geq 0.
\end{equation}
Moreover, we have, again by Hardy's inequality,
\begin{equation}
\begin{split}
  \<f,Af\> &= \|\nabla f\|^2_\L{2} + \<f,Vf\> 
  \geq \frac{1}{4}\int \frac{|f(x)|^2}{|x|^2}\;d^3x + \int |V(x)|\, |f(x)|^2\;d^3x \\
  &\geq \int \left[\frac{1}{4|x|^2}-\frac{V_0}{|x|^2}\right] |f(x)|^2\;d^3x
  \geq \delta \int \frac{|f(x)|^2}{|x|^2}\;d^3x
\end{split}
\end{equation}
where $\delta:=\frac{1}{4}-V_0>0$. Then, $\<f,Af\>=0$ implies $\int \frac{|f(x)|^2}{|x|^2}\;d^3x=0$ and hence $|f(x)|=0$ (a.e. on $\R^3$). This shows that $A$ is positive definite. 
Then, according to \cite{Wouk}, $A$ has a unique positive definite self-adjoint square root $B:=A^{1/2}$ with $\Dom(B)=\{f\in \L{2}: Bf\in \L{2}\}\supset \Dom(A)$. In spirit of the spectral theorem we can define positive self-adjoint $A^{s/2}$ for any $s\geq 0$.

We recall that the homogeneous $\H{s}$ and inhomogeneous $H^{s}$ Sobolev spaces are defined as completions of $C_0^\infty$ w.r.t the norms
\begin{equation}
  \|h\|_\H{s} := \|(-\lap)^{s/2} h\|_\L{2} \qquad\text{and}\qquad
  \|h\|_{H^s}^2 := \sum_{k=0}^s \|h\|_\H{k}^2 = \sum_{k=0}^s \|(-\lap)^{k/2} h\|_\L{2}^2,
\end{equation}
respectively, and have the property $\H{0}\cap\H{s}=H^s$. 
In order to use the standard construction of a unitary evolution given by the self-adjoint generator $A$ we define \textit{perturbed} homogeneous $\Hd{s}$ and inhomogeneous $\Hh{s}$ Sobolev spaces as completions of $C_0^\infty$ w.r.t the norms (in case $A$ is positive but not positive definite $\|\cdot\|_\Hd{s}$ are only seminorms)
\begin{equation}
  \|h\|_\Hd{s} := \|A^{s/2} h\|_\L{2} \qquad\text{and}\qquad
  \|h\|_\Hh{s}^2 := \sum_{k=0}^s \|h\|_\Hd{k}^2 = \sum_{k=0}^s \|A^{k/2} h\|_\L{2}^2.
\end{equation}
for $s\in\N$. In our situation $A$ is positive definite, so $\|\cdot\|_\Hd{s}$ and $\|\cdot\|_\Hh{s}$ are norms and $\Hd{s}$ and $\Hh{s}$ are (complete) Banach spaces.

Now we relate $\Hd{1}$ to energy. Consider any $f\in\COinf$ for that we can partially integrate
\begin{equation}
\begin{split}
  \|f\|^2_\Hd{1} &= \<A^{1/2}f,A^{1/2}f\> = \<f,A f\> \\
  &= \<f,-\lap f\> + \<f,Vf\> = \<\nabla f,\nabla f\> + \<f,Vf\> =: \|f\|^2_{Y_E}.
\end{split}
\end{equation}
The last expression defines what we will call ``partial energy'' norm (its norm properties are inherited from $\|\cdot\|_\Hd{1}$). If we define the ``partial energy'' space $Y_E$ by completion of $\COinf$ w.r.t. this norm then the spaces $Y_E$ and $\Hd{1}$ are identical, because both are defined as completions of $\COinf$ w.r.t. to the norms which coincide\footnote{More precisely, the completions are defined by Cauchy sequences in $\COinf$ w.r.t. the given norms. Since both norms are equal on $\COinf$, the Cauchy sequences w.r.t. to the two norms are identical.} on the whole $\COinf$. Hence, $Y_E\equiv \Hd{1}$
(see also \cite{Burq} for similar considerations). 

We can now define the energy space\footnote{We choose Cartesian products instead of direct sums in order to avoid confusion with norms. A direct sum of Banach spaces induces a norm and a direct sum of Hilbert spaces induces an inner product, which in turn induces a norm, but these two norms are different. Since we treat the same spaces as Banach and Hilbert we use rather Cartesian products and specify the norms explicitly every time we define a new space.} as $X_E:=Y_E\times \L{2}=\Hd{1}\times \L{2}$ or equivalently as a completion of $\COinf$ w.r.t. the norm 
\begin{equation} \label{E-HL}
  \|(f,g)\|^2_{X_E} := \|\nabla f\|^2_\L{2} + \<f,Vf\> + \|g\|_\L{2}^2 =
  \|f\|^2_\Hd{1} + \|g\|_\L{2}^2.
\end{equation}
$X_E$ is a Hilbert space with the scalar product
\begin{equation}
  \<(f_1,g_1)|(f_2,g_2)\>_{X_E} := \<A^{1/2}f_1,A^{1/2}f_2\> + \<g_1,g_2\>, \qquad (f_i,g_i)\in \Hd{1}\times\L{2},
\end{equation}
what follows from the fact that every Banach space $\Hd{s}$ can be made to a  Hilbert space equipped with the scalar product
\begin{equation}
  \<\cdot,\cdot\>_\Hd{s} := \<A^{s/2} \cdot,A^{s/2} \cdot\>_\L{2}.
\end{equation}
We enhance the facts that $A$ and $B$ are self-adjoint on $\L{2}$ with the domains $\Dom(A)=\Hh{2}\subset\L{2}$ and $\Dom(B)=\Hh{1}\subset\L{2}$, respectively, by 
the observation that $B$ defined on $\Hd{1}$ is self-adjoint with $\Dom_\Hd{1}(B)=\Hd{1}\cap\Hd{2}\subset\Hd{1}$.

For more flexibility in the applications we also relate the homogeneous and inhomogeneous Sobolev spaces. 
For $s=0$ we trivially have $\Hd{0}=\H{0}$. From \refa{fVf-Hardy} and \refa{Df-fVf-Df} it follows
\begin{equation}
  0 \leq (1-4V_0) \|\nabla h\|_\L{2}^2 \leq 
  \|\nabla h\|_\L{2}^2 + \<h,V h\>  \leq
  (1+4V_0) \|\nabla h\|_\L{2}^2,
\end{equation}
what means that there exist constants $C_1,C_2>0$ such that $C_1 \|h\|_\H{1} \leq \|h\|_{Y_E} \leq C_2 \|h\|_\H{1}$, hence the norms in $\H{1}$ and $Y_E$ are equivalent and $\H{1}=Y_E$ as sets. Since $Y_E \equiv \Hd{1}$ we have $\H{1}=\Hd{1}$ as sets with equivalent norms. The same is also true for the inhomogeneous spaces, namely $H^{1}=\Hh{1}$ as sets with equivalent norms.

Furthermore, since $\|Vh\|_\L{2}\leq V_0 \|h\|_\L{2}$ we have on the one hand that $h\in \L{2} \cap \H{2}$ implies $\|Ah\|_\L{2} \leq \|-\lap h\|_\L{2} + V_0 \|h\|_\L{2}<\infty$ and hence $h\in \Hd{2}$. On the other hand, $h\in \L{2} \cap \Hd{2}$ implies $\|-\lap h\|_\L{2} \leq \|A h\|_\L{2} + V_0 \|h\|_\L{2}<\infty$ and hence $h\in \H{2}$. That gives $H^2=\Hh{2}$.

Now, we are ready to formulate and prove global existence, uniqueness and energy conservation for solutions of the wave equation in the energy space $X_E$.
\renewcommand{\theTheoremA}{1'}
\begin{TheoremA} \label{Th-EnergyConserv}
  Assume the potential $V$ satisfies the bound \refa{V-bound-App} with $V_0<\frac{1}{4}$.
  
  a) For all initial data $(f,g)\in \Hd{1}\times\L{2}$ there exists a unique weak solution $u\in\C^0(\Rb_+,\Hd{1})\cap\C^1(\Rb_+,\L{2})$ of the wave equation \refa{wave-eq}. Moreover, the energy defined by the norm in $X_E$ is constant in time
\begin{equation} \label{th-E}
\begin{split}
  \|(u(t),\d_t u(t))\|^2_{X_E} &= \<\nabla u(t), \nabla u(t)\> + \<u(t), V u(t)\> + \<\d_t u(t), \d_t u(t)\>\\
  &= \<\nabla f, \nabla f\> + \<f, V f\> + \<g, g\>.
\end{split}
\end{equation}

b) Initial data $(f,g)\in\H{1}\times\L{2}$ or $H^{1}\times\L{2}$ give rise to a unique solution $u\in\C^0(\Rb_+,\H{1})\cap\C^1(\Rb_+,\L{2})$ or $\C^0(\Rb_+,H^{1})\cap\C^1(\Rb_+,\L{2})$, respectively, but the corresponding norms are not conserved in evolution.

\end{TheoremA}
\begin{proof} a)
In order to prove existence of a unique (weak) solution in $X_E=\Hd{1}\times\L{2}$ we take advantage of the functional analysis in Hilbert spaces. We show that the evolution in $X_E$ is generated by a self-adjoint operator $M$ and use the famous theorem of Stone to show that there exists a unique strongly continuous group of unitary operators $W(t)$ defining the evolution of $u$. Unitarity of $W(t)$ guarantees conservation of the norm in $X$, which represents the energy of $u(t)$.

First, we write the wave equation \refa{wave-eq} as
\begin{equation} \label{wave-eq-A}
  - \dt[2] u = A u = (- \lap + V(x)) u.
\end{equation}
$A$ with $\Dom(A)=\Hh{2}$ is self-adjoint on the Hilbert space $Z:=\L{2}$ and positive definite when the bound \refa{V-bound-App} is satisfied with $V_0<\frac{1}{4}$ (cf. \refa{Df-fVf-Df}). Its unique positive definite square root $B:=A^{1/2}$ is self-adjoint with $\Dom_Z(B)=\Hh{1}\subset Z$ on the same Hilbert space $Z=\L{2}$. On the Hilbert space $Y:=\Hd{1}$ with the scalar product $\<\cdot,\cdot\>_Y:=\<B\cdot,B\cdot\>_Z$ the operator $B$ is again self-adjoint with $\Dom_Y(B)=\Hd{1}\cap\Hd{2}\subset Y$.

Next, the wave equation \refa{wave-eq-A} can be written as a system of first order differential equations in time on $X_{ZZ}:=\Dom_Z(B)\times Z=\Hh{1}\times \L{2}$ with $\<(f_1,g_1),(f_2,g_2)\>_{X_{ZZ}} := \<f_1,f_2\>_Z + \<B f_1,B f_2\>_Z + \<g_1,g_2\>_Z$
\begin{equation}
  \dt U(t) = -i M U(t),\qquad 
  U(0) =  \left( \begin{array}{c} f \\ g \end{array} \right)
\end{equation}
with
\begin{equation}
  U(t):=\left( \begin{array}{c} u(t) \\ \dot{u}(t) \end{array} \right)\in X_{ZZ}
  \qquad\text{and}\qquad
  M:=i\left( \begin{array}{rr} 0\quad & 1 \\ -A\quad & 0 \end{array} \right).
\end{equation}
$M$ generates a strongly continuous semigroup $W_{ZZ}(t)$ on $X_{ZZ}$ which solves the above initial value problem, but $M$ is not self-adjoint on $X_{ZZ}$ (not even symmetric) and hence $W_{ZZ}(t)$ is not unitary on $X_{ZZ}$.
Therefore, we consider now a completion $X_{YZ}$ of $X_{ZZ}$ in the norm $\|(f_1,g_1)\|^2_{X_{YZ}} := \|B f_1\|_Z^2 + \|g_1\|_Z^2 = \|f_1\|_Y^2 + \|g_1\|_Z^2$ equipped with the scalar product $\<(f_1,g_1),(f_2,g_2)\>_{X_{YZ}} := \<B f_1,B f_2\>_Z + \<g_1,g_2\>_Z = \<f_1,f_2\>_Y + \<g_1,g_2\>_Z$. It turns out that $X_{YZ}=Y\times Z=\Hd{1}\times \L{2}=X_E$.
Then the closure of $M$ with $\Dom(M)=\Dom(A)\times\Dom_Z(B)=\Hh{2}\times\Hh{1}$ is self-adjoint on $X_E$ (see \cite[Lem.7.7]{Goldstein} for a general proof or \cite{Thoe-WavPot} for the case $V\geq 0$ treated more explicitly) and generates a strongly continuous unitary group $W(t)$ on $X_E$ given by the unique extension of
\begin{equation}
  W(t):=\left( \begin{array}{rr} \cos(tB)\quad & B^{-1}\sin(tB) 
  \\ -B\sin(tB)\quad & \cos(tB) \end{array} \right)
\end{equation}
to $X_E$ (see also \cite[sec. X.13]{RS-II} or \cite[sec. 6.2]{Taylor-I} for similar considerations). Such $W(t)$ defines a strongly continuous evolution in $X_E$
\begin{equation}
  U(t)=W(t) U(0) = W(t) \left( \begin{array}{c} f \\ g \end{array} \right).
\end{equation}
Thus, we have $U=(u,\dot{u})\in\C^0(\Rb_+,\Hd{1}\times\L{2})$ or, in other words, $u\in\C^0(\Rb_+,\Hd{1})\cap\C^1(\Rb_+,\L{2})$.

From the unitarity of $W(t)$ we get $\|U(t)\|_{X_E} = \|U(0)\|_{X_E}$ what expressed in $u(t)$ reads
\begin{equation}
  \|u(t)\|_\Hd{1}^2 + \|\dot{u}(t)\|_\L{2}^2 = \|f\|_\Hd{1}^2 + \|g\|_\L{2}^2.
\end{equation}
Since the above norm in $\Hd{1}\times\L{2}$ is equal to the energy norm $\|\cdot\|_{X_E}^2$
we get the energy conservation \refa{th-E}.

b)
Conservation of energy in $X_E$ implies that the solution also belongs to the free energy space $X_0=\H{1}\times\L{2}$
\begin{equation}
  E_0[u(t)] = \|\nabla u(t)\|_\L{2}^2+\|\d_t u(t)\|_\L{2}^2 \leq C_1 E[u(t)] = C_1 E[u(0)] < \infty
\end{equation}
with $C_1:=(1-4V_0)^{-1}>1$ what follows from \refa{Df-fVf-Df} and $V_0<\frac{1}{4}$ (for $V\geq 0$ the same holds with $C_1=1$). Moreover, by \refa{fVf-Hardy},
\begin{equation}
  E_0[u(t)] \leq C_1 E[u(0)] \leq C_2 E_0[u(0)].
\end{equation}
Summarizing this, there is a continuous (but not unitary) evolution $(f,g)\in \H{1}\times \L{2} \ra (u(t),\d_t u(t))\in \H{1}\times \L{2}$ for all $t>0$.

In case when additionally $f\in\L{2}$ we get $u(t)\in\L{2}$ for all $t\in\Rb_+$, because
\begin{equation}
\begin{split}
  \|u(t)\|_\L{2} &= \left\|u(0) + \int_0^t \d_t u(t') dt'\right\|_\L{2} 
  \leq \|f\|_\L{2} + \int_0^t \|\d_t u(t')\|_\L{2} dt' \\
  &\leq \|f\|_\L{2} + \int_0^t \sqrt{E[u(t)]} dt' 
  \leq \|f\|_\L{2} + \sqrt{E[u(0)]}\; t < \infty \qquad \forall t>0,
\end{split}
\end{equation}
and a continuous evolution $(f,g)\in H^1\times \L{2} \ra
(u(t),\d_t u(t))\in H^1\times \L{2}$ for all $t>0$.
\end{proof}

\Proof[Theorem \ref{Th-Energy}]

\begin{proof}
  In \refa{f-g-L2}-\refa{E-finite} we have shown that the assumptions of Theorem \ref{Th-decay} or \ref{Th-Ldecay} imply finiteness of the energy of initial data $E[u(0)] <\infty$. Since by \refa{E-HL}
\begin{equation}
  \|f\|_\Hd{1}^2 + \|g\|_\L{2}^2 = E[u(0)] < \infty,
\end{equation}
we have $(f,g)\in\Hd{1}\times\L{2}$. The condition $V_0<C^{-1}_{p,k}$ implies $V_0<\frac{1}{4}$ for any $p>1, k>2$. So we can apply Theorem \ref{Th-EnergyConserv} a) and obtain a unique solution $u\in\C^0(\Rb_+,\Hd{1})\cap\C^1(\Rb_+,\L{2})$ with conserved energy 
\begin{equation}
  E[u(t)]=\|(u(t),\d_t u(t))\|^2_{X_E} = \<\nabla f, \nabla f\> + \<f, V f\> + \<g, g\> = E[u(0)].
\end{equation}
Since the assumptions of Theorem \ref{Th-decay} or \ref{Th-Ldecay} imply also $(f,g)\in H^1\times\L{2}$ we have by Theorem \ref{Th-EnergyConserv} b) that $u\in \C^0(\Rb_+,H^{1})\cap\C^1(\Rb_+,\L{2})$.

This gives $u(t), \nabla u(t), \d_t u(t) \in \L{2}$ for all $t\in\Rb_+$ what implies $u(t), \nabla u(t), \d_t u(t) \in \Lloc$ for all $t\in\Rb_+$, i.e. existence of the first derivatives in the weak sense, because for any compact set $\Omega\subset\R^3$ and $h\equiv u(t), \nabla u(t)$ or $\d_t u(t)$ H{\"o}lder's inequality gives
\begin{equation}
  \int_\Omega |h|\, d\mu \leq \left( \int_\Omega 1\,d\mu \right)^{1/2} \cdot 
  \left( \int_\Omega |h|^2\,d\mu \right)^{1/2} 
  \leq \mu(\Omega)^{1/2} \cdot \|h\|_{\L{2}(\Omega)} < \infty.
\end{equation}
The last step is to show that such constructed $u\in \C^0(\Rb_+,H^{1})\cap\C^1(\Rb_+,\L{2})$ really solves the equation \refa{wave-eq}. Indeed, it solves the equation in the distributional sense (see \cite[Th. 3.2]{Sogge-book} for a more detailed discussion). Since additionally $u\in\Lloc$ it solves \refa{wave-eq} in the weak sense \refa{weak-sol-pot}.
\end{proof}


\section{Proofs of the estimates} \label{Sec-Proofs}

\Proof[Lemma \ref{Lem-int}]
\begin{proof}
For $t>0$ we have
\begin{equation}
  I =\frac{1}{4\pi}\int_{S(x,t)} \frac{d\sigma(y)}{f(|y|)}
  = \frac{1}{4\pi} \int_{S(0,1)} \frac{t^2 d\sigma(\omega)}{f(|x+t\omega|)}  
  = \frac{t^2}{4\pi} \int_{S(0,1)} \frac{d\sigma(\omega)}{f(|x+t\omega|)}  
\end{equation}
Using the fact that $|\omega|=1$ and introducing the angle $\theta$ between $x$ and $\omega$ we get
\begin{equation}
  |x+t\omega|^2=|x|^2+t^2|\omega|^2+2t(x\omega) = |x|^2+t^2+2t|x|\cos\theta
\end{equation}
hence in polar coordinates $(\theta, \phi)$
\begin{equation}
\begin{split}
  I&=\frac{t^2}{4\pi}\int_0^\pi d\theta \sin\theta \int_0^{2\pi} d\phi   
  \frac{1}{f(\sqrt{|x|^2+t^2+2t|x|\cos\theta})} \\
  &= \frac{t^2}{2} \int_{-1}^{+1} d\chi \frac{1}{f(\sqrt{|x|^2+t^2+2t|x|\chi})},
\end{split}  
\end{equation}
where we substituted $\chi:=\cos\theta$. Now, changing variables $\la^2:=|x|^2+t^2+2t|x|\chi$ and $\la d\la=t|x|d\chi$ we get
\begin{equation}
  I = \frac{t^2}{2} \int_{|t-|x||}^{t+|x|} \frac{\la d\la}{t|x|f(\la)}
  = \frac{t}{2|x|} \int_{|t-|x||}^{t+|x|} \frac{\la d\la}{f(\la)}.
\end{equation}
Next, choosing $f(\la)=\norm{x}^p=(1+\la)^p$ and $p>2$ we have
\begin{equation} 
  I=\frac{t}{2|x|} \int_{|t-|x||}^{t+|x|} \frac{\la d\la}{(1+\la)^p}.
\end{equation}
This simple integral can be evaluated exactly for $p-2>0$ and then estimated
\begin{equation} \label{est-int}
\begin{split}
  I&=\frac{t}{2|x|} \frac{-1}{(p-2)} 
   \left.\frac{1}{(1+\la)^{p-2}} \left[ 1-\frac{p-2}{(p-1)(1+\la)}\right] \right|_{|t-|x||}^{|t+|x||} \\
  &= \frac{t}{2|x|} \left[\frac{1}{(p-2) (1+|t-|x||)^{p-2}} - \frac{1}{(p-2) (1+|t+|x||)^{p-2}}
   \right. \\
  &+ \underbrace{\frac{1}{(p-1) (1+|t+|x||)^{p-1}} - \frac{1}{(p-1) (1+|t-|x||)^{p-1}}}_{\leq 0}\bigg] \\ 
  &\leq \frac{t}{2(p-2)|x|} \frac{1}{\norm{t-|x|}^{p-2}}.
\end{split}  
\end{equation}
It gives the estimate \refa{lem1-est} with $c_p:=1/2(p-2)$.

We proceed further in two ways: for $q>2$, on the one hand we have from \refa{est-int}
\begin{equation} \label{est-A}
  \frac{1}{4\pi t} \int_{S(x,t)} \frac{d\sigma(y)}{\norm{y}^q}
  \leq \frac{c_p}{|x| \norm{t-|x|}^{q-2}}.
\end{equation}
and the other hand, taking the supremum out of the integral, we find
\begin{equation} \label{est-B}
  \frac{1}{4\pi t} \int_{S(x,t)} \frac{d\sigma(y)}{\norm{y}^q}
  \leq \frac{1}{4\pi t} \left(\sup_{|y-x|=t}\frac{1}{\norm{y}^q}\right) \int_{S(x,t)} d\sigma(y)
  = \frac{1}{4\pi t} \frac{4\pi t^2}{\norm{t-|x|}^q}
  = \frac{t}{\norm{t-|x|}^q}.
\end{equation}
Combining both inequalities \refa{est-A}, \refa{est-B} for $q=p>2$ we get
\begin{equation}
  \frac{1}{4\pi t} \int_{S(x,t)} \frac{d\sigma(y)}{\norm{y}^p}
  \leq \frac{1}{\norm{t-|x|}^{p-2}} 
  \min \left(\frac{2\,c_p}{2|x|}, \frac{t}{\norm{t-|x|}^2}\right).
\end{equation}
By little algebra it can be shown that
\begin{alignat}{3}
  \frac{2\,c_p}{2|x|} &\leq \frac{9\,c_p}{\norm{t+|x|}}&\qquad&
  \text{for}\quad&|x|&\geq\frac{t}{4},\; |x|\geq \frac{1}{4}\\
  \frac{t}{\<t-|x|\>^2} &\leq \frac{4}{\norm{t+|x|}}&&
  \text{for}&|x|&\leq \frac{t}{4},\; t\geq 1\\
  \frac{t}{\<t-|x|\>^2} &\leq \frac{\frac{9}{4}}{\norm{t+|x|}}&&
  \text{for}&|x|&\leq \frac{1}{4},\; t\leq 1.
\end{alignat}
Therefore, for $p>2$
\begin{equation} 
  \boxed{\frac{1}{4\pi t} \int_{S(x,t)} \frac{d\sigma(y)}{\norm{y}^p}
  \leq \frac{C^{(1)}_p}{\norm{t+|x|}\norm{t-|x|}^{p-2}}}
\end{equation}
where
\begin{equation}
  C^{(1)}_p := \max\left(9\,c_p, 4, \frac{9}{4}\right)= 
  \max \left(\frac{9}{2(p-2)}, 4\right).
\end{equation}
Now, combining inequalities \refa{est-A}, \refa{est-B} for $q=p-1>2$ we get
\begin{equation}
  \frac{1}{4\pi t^2} \int_{S(x,t)} \frac{d\sigma(y)}{\norm{y}^{p-1}}
  \leq \frac{1}{\norm{t-|x|}^{p-3}} 
  \min \left(\frac{2c_{p-1}}{2|x|t}, \frac{1}{\norm{t-|x|}^2}\right).
\end{equation}
By some more algebra it can be shown that
\begin{alignat}{2}
  &\text{for } \frac{1}{2} \leq \frac{|x|}{t} \leq 2\text{ and }|x|,t\geq 1:\qquad &
  \frac{2\,c_{p-1}}{2|x|t} &\leq \frac{6\,c_{p-1}}{\norm{t-|x|}\norm{t+|x|}}, \\
  &\text{for } \frac{|x|}{t} \geq 2\text{ or } \frac{|x|}{t} \leq \frac{1}{2}:\quad &
  \frac{1}{\norm{t-|x|}^2} &\leq \frac{5}{\norm{t-|x|}\norm{t+|x|}}, \\
  &\text{for } |x|,t\leq 2:&
  \frac{1}{\norm{t-|x|}^2} &\leq \frac{5}{\norm{t-|x|}\norm{t+|x|}}.
\end{alignat}
So we have
\begin{equation}
  \min \left(\frac{2c_{p-1}}{2|x|t}, \frac{1}{\norm{t-|x|}^2}\right) \leq 
  \frac{\max(6\,c_{p-1},5,5)}{\norm{t-|x|}\norm{t+|x|}}.
\end{equation}
So we finally get for $p>3$
\begin{equation} 
  \boxed{\frac{1}{4\pi t^2} \int_{S(x,t)} \frac{d\sigma(y)}{\norm{y}^{p-1}}
  \leq \frac{C^{(2)}_p}{\norm{t+|x|}\norm{t-|x|}^{p-2}} }
\end{equation}
where
\begin{equation}
  C^{(2)}_p := \max (6\,c_{p-1},5) =  \max \left(\frac{3}{p-1}, 5\right).
\end{equation}
\end{proof}

\Proof[Lemma \ref{Lem-int-cone}]
For completeness we cite the proof after Georgiev \textit{et al.} \cite{Georg-H-K}. 
\begin{proof}
For $t>0$ and the continuous integrand in $I$ we use a version of Fubini's theorem and split the integration over the cone $K(x,t)$ into integrations over time and over spheres 
\begin{equation}
  I = \frac{1}{4\pi}\int_0^t ds \int_{S(x,t-s)} d\sigma(y)\; 
  \frac{1}{(t-s) \norm{y}^q \norm{s+|y|} \norm{s-|y|}^{p-1}}.
\end{equation}
For integration over the spheres $S(x,t-s)$ we use lemma \ref{Lem-int} and obtain
\begin{equation}
\begin{split}
  I &= \frac{1}{2|x|} \int_0^t ds \int_{|t-s-|x||}^{t-s+|x|} d\la\; 
  \frac{\la}{\norm{\la}^q \norm{s+\la} \norm{s-\la}^{p-1}}\\
  &= \frac{1}{2|x|} \int_0^t ds \int_{|t-s-|x||}^{t-s+|x|} d\la\; 
  \frac{1}{(1+\la)^{q-1} (1+s+\la) (1+|s-\la|)^{p-1}}.
\end{split}
\end{equation}
One can estimate the integrand for $q\geq p$ 
\begin{equation}
  \frac{1}{(1+\la)^{q-1} (1+s+\la) (1+|s-\la|)^{p-1}} \leq 
  \frac{3^{q-1}}{(1+s+\la)^{p}} \left(\frac{1}{(1+|s-\la|)^{q-1}} + \frac{1}{(1+\la)^{q-1}}\right)
\end{equation}
because for $0\leq s \leq 2\la$ it is smaller than the first term and for $s\geq 2\la$ smaller than the second one (see \cite{Georg-H-K} or \cite{Strauss-T} for more details). Now, changing variables $\alpha := \la+s$ and $\beta := \la-s$, we get for $q>2$
\begin{equation}
\begin{split}
  I&\leq \frac{3^{q-1}}{2|x|} \int_{|t-|x||}^{t+|x|} \frac{d\alpha}{(1+\alpha)^{p}}
  \int_{-\infty}^{+\infty} d\beta \left[ \frac{1}{(1+|\beta|)^{q-1}} + \frac{1}{\left(1+\frac{|\alpha+\beta|}{2}\right)^{q-1}} \right] \\
  &\leq \frac{3^{q-1}}{2|x|} \frac{2^{q-1}}{q-2}\underbrace{(1+2^{1-q})}_{\leq 3/2} \int_{|t-|x||}^{t+|x|} \frac{d\alpha}{(1+\alpha)^{p}}.
\end{split}
\end{equation}
The last integral can be estimated either by taking supremum of the integrand or by integrating it explicitly
\begin{equation}
  J:=\frac{1}{2|x|}\int_{|t-|x||}^{t+|x|} \frac{d\alpha}{(1+\alpha)^{p}} 
  \leq \left\{ \begin{array}{l}
    (1+|t-|x||)^{-p-1} \\
    (2(p-1)|x|)^{-1} (1+|t-|x||)^{-p}
  \end{array} \right.
  = \frac{1}{\norm{t-|x|}^{p-1}} \left\{ \begin{array}{l}
    \norm{t-|x|}^{-1} \\
    (2(p-1)|x|)^{-1}
  \end{array} \right. .
\end{equation}
It can be easily shown that
\begin{alignat}{3}
  \frac{1}{2p|x|} &\leq \frac{2/(p-1)}{\norm{t+|x|}}&\qquad&
  \text{for}\quad&2|x|&\geq t,\; 2|x|\geq 1\\
  \frac{1}{\<t-|x|\>} &\leq \frac{3}{\norm{t+|x|}}&&
  \text{for}&2|x|&\leq t,\; t\geq 1\\
  \frac{1}{\<t-|x|\>} &\leq \frac{3}{\norm{t+|x|}}&&
  \text{for}&t,|x|&\leq 1
\end{alignat}
and hence
\begin{equation}
  J\leq \frac{\max(2/(p-1),3)}{\norm{t+|x|}\norm{t-|x|}^{p-1}}.
\end{equation}
Finally, we obtain
\begin{equation}
  I\leq  \frac{C_{p,q}}{\norm{t+|x|}\norm{t-|x|}^{p-1}}
\end{equation}
with
\begin{equation}
  C_{p,q}:=\frac{3}{2} \frac{6^{q-1}}{(q-2)} \max(2/(p-1),3). \tag*{\qedhere}
\end{equation}
\end{proof}

\Proof[Lemma \ref{Lem-weak-sol}]
\begin{proof}
For any distributions $f,g\in\D(\R^3)$ and $F\in\D(\RTXa)$ with support on $\RTXb$ the representation formula \refa{lem-Duhamel-conv} can be written as
\begin{equation}
\begin{split}
  v(t,x)&=\d_t \int_{\R} ds \int_{\R^3} d^3y \frac{\delta(t-s-|x-y|)}{4\pi (t-s)} \delta(s)f(y) 
  + \int_{\R} ds \int_{\R^3} d^3y \frac{\delta(t-s-|x-y|)}{4\pi (t-s)} \delta(s)g(y) \\
  &+ \int_{\R} ds \int_{\R^3} d^3y\; \frac{\delta(t-s-|x-y|)}{4\pi (t-s)} \theta(s) F(s,y)\\
  &=\d_t \left[\frac{\delta(\cdot_t-|\cdot_x|)}{4\pi \cdot_t}\ast\delta(\cdot_t)f(\cdot_x)\right](t,x)
  +\left[ \frac{\delta(\cdot_t-|\cdot_x|)}{4\pi \cdot_t} \ast \delta(\cdot_t) g(\cdot_x) \right](t,x)\\
  &+ \left[ \frac{\delta(\cdot_t-|\cdot_x|)}{4\pi \cdot_t} \ast \theta(\cdot_t)F \right](t,x)\\
  &= \left[ \frac{\delta(\cdot_t-|\cdot_x|)}{4\pi \cdot_t} \ast \left[ 
  \delta'(\cdot_t)f(\cdot_x) + \delta(\cdot_t) g(\cdot_x) + \theta(\cdot_t)F\right] \right](t,x),
\end{split}
\end{equation}
i.e. a sum of $\RTXa$-convolutions of distributions $\delta(t)f(x), \delta(t)g(x), \theta(t)F(t,x)\in\D(\RTXa)$ with $\delta(t-|x|)/(4\pi t)\in\D(\RTXa)$, all supported on $\RTXb$. 
The formula defines again a distribution $v\in\D(\RTXa)$ \cite{Rudin} supported on $\RTXb$.
In order to show that $v$ solves weakly the wave equation \refa{wave-eq-dist}, i.e. $v$ satisfies \refa{weak-sol-lem}, we transform the l.h.s. of \refa{weak-sol-lem} for any test function $\varphi\in\C_0^\infty(\R^{1+3})$
\begin{equation} 
\begin{split}
  &\int dt \int d^3x\; v(t,x)\; \Box \varphi(t,x) =\\ 
  &=\int dt \int d^3x\; \left[ \frac{\delta(\cdot_t-|\cdot_x|)}{4\pi \cdot_t} \ast \left[ 
  \delta'(\cdot_t)f(\cdot_x)+\delta(\cdot_t)g(\cdot_x)+\theta(\cdot_t)F\right]\right](t,x)\;\Box \varphi(t,x)\\
  &=\int dt \int d^3x\; \Box \left[ \frac{\delta(\cdot_t-|\cdot_x|)}{4\pi \cdot_t} \ast \left[ 
\delta'(\cdot_t)f(\cdot_x)+\delta(\cdot_t)g(\cdot_x)+\theta(\cdot_t)F\right]\right](t,x)\;\varphi(t,x)\\
  &=\int dt \int d^3x\; \left[\Box\left(\frac{\delta(\cdot_t-|\cdot_x|)}{4\pi \cdot_t}\right) \ast
\left[\delta'(\cdot_t)f(\cdot_x)+\delta(\cdot_t)g(\cdot_x)+\theta(\cdot_t)F\right]\right](t,x)\;\varphi(t,x)
\end{split}
\end{equation}
where we have first used the definition of differentiation of distributions (corresponding to partial integration) and then the theorem on differentiation of convolutions \cite{Rudin}. Now, with the well-known distributional identity
\begin{equation}
  \Box \left(\frac{\delta(t-|x|)}{4\pi t}\right) = \delta(t) \delta(x),
\end{equation}
we arrive at
\begin{equation} \label{weak-sol-proof}
\begin{split}
  &\int dt \int d^3x\; v(t,x)\; \Box \varphi(t,x) =\\ 
  &=\int dt \int d^3x\; \left[\delta'(t)f(x) + \delta(t) g(x) + \theta(t)F(t,x)\right]\;\varphi(t,x)\\
  &= - \int d^3x\; f(x)\;\d_t \varphi(0,x) + \int d^3x\; g(x) \; \varphi(0,x)
  + \int_{\Rb_+} dt \int d^3x\; F(t,x)\;\varphi(t,x).
\end{split}
\end{equation}
In case when $f,g$ and $F$ are ordinary functions or in $\Lloc$, then they are also (regular) distributions with, in general, distributional derivatives. For them the statements above remain untouched. If additionally $v\in\Lloc$ is defined almost everywhere as a function on $\RTXb$ by the Duhamel's formula \refa{lem-Duhamel}, which is equivalent to \refa{lem-Duhamel-conv} with only alternative notation for the measures ($\delta(...)\,d^3y \ra d\sigma(y)$), then it also solves weakly the same wave equation, i.e. its derivatives exist in the distributional sense and $v\in\Lloc$ satisfies \refa{weak-sol-proof}.
\end{proof}

\Proof[Lemma \ref{Lem-free_wave_eq}]
\begin{proof}
Let $v(t,x)$ for $t>0$, $x\in\R^3$ be given by the formula
\begin{equation} \label{Duhamel}
\begin{split}
  v(t,x)&=\frac{1}{4\pi}\int_{S(x,t)} \frac{g(y)}{t} d\sigma(y)
  +\frac{1}{4\pi}\int_{S(x,t)} \frac{(y-x)\cdot\nabla f(y)+f(y)}{t^2} d\sigma(y)\\
  &= \frac{t}{4\pi} \int_{S(0,1)} g(x+t\omega) \; d\sigma(\omega) + 
  \frac{t}{4\pi}\int_{S(0,1)} \omega\cdot\nabla f(x+t\omega) \; d\sigma(\omega)\\
  &+ \frac{1}{4\pi} \int_{S(0,1)} f(x+t\omega)\; d\sigma(\omega).
\end{split}
\end{equation}
Since $g,f,\nabla f\in\C^0$ these integrals exist for all $(t,x)\in\RTX$ and are finite. 
Moreover, since all three above integrands are continuous functions of $(t,x)$ and are uniformly bounded by $g_0, f_1, f_0\in L^1(S(0,1))$ for all $(t,x)$, from continuity of Lebesgue integrals, we obtain that $v$ is also continuous in $(t,x)\in\RTX$. The first two integrals are bounded by $4\pi g_0$ and $4\pi f_1$, respectively, hence $\lim_{t\ra 0^+} v(t,x) = f(x) = v(0,x)$, so we have $v\in\C^0(\RTXb)$. Then also $v\in\Lloc$ and from lemma \ref{Lem-weak-sol} it follows that $v$ solves weakly the wave equation \refa{lem-wave-eq}.

Therefore, we are able to apply the pointwise estimates
\begin{equation} \label{est-2int}
\begin{split}
  |v(t,x)| &\leq \frac{1}{4\pi} \int_{S(x,t)} \left(\frac{|g(y)|}{t} 
  + \frac{|\nabla f(y)|}{t} + \frac{|f(y)|}{t^2} \right)  d\sigma(y) \\
  &\leq \frac{\|\norm{\cdot}^m g\|_\infty + \|\norm{\cdot}^m \nabla f\|_\infty}{4\pi t} 
  \int_{S(x,t)} \frac{d\sigma(y)}{\norm{y}^m}
  + \frac{\|\norm{\cdot}^{m-1} f\|_\infty}{4\pi t^2} \int_{S(x,t)} \frac{d\sigma(y)}{\norm{y}^{m-1}}\\
  &= \frac{g_0 + f_1}{4\pi t} \int_{S(x,t)} \frac{d\sigma(y)}{\norm{y}^m}
  +  \frac{f_0}{4\pi t^2} \int_{S(x,t)} \frac{d\sigma(y)}{\norm{y}^{m-1}}.
\end{split}
\end{equation}
From lemma \ref{Lem-int}, eq. \refa{est-int1} and \refa{est-int2}, for $m>3$, we finally obtain
\begin{equation}
  |v(t,x)| \leq \frac{C_m\cdot(g_0 + f_1 + f_0)}{\norm{t+|x|}\norm{t-|x|}^{m-2}} \qquad \forall (t,x)\in\RTX
\end{equation}
with
\begin{equation} \label{Cp}
  C_m:= \max(C^{(1)}_m,C^{(2)}_m) = \max \left(\frac{9}{2(m-2)}, 5\right). 
\end{equation}
The inclusion of the boundary $t=0$ is trivial, because from the bounds on the initial data we have
\begin{equation}
  |v(0,x)| = |f(x)| \leq \frac{f_0}{\norm{|x|}^{m-1}} \leq  
  \frac{C_m\cdot(g_0 + f_1 + f_0)}{\norm{0+|x|}\norm{0-|x|}^{m-2}} \qquad \forall x\in\R^3.
\end{equation}
\end{proof}

\Proof[Lemma \ref{Lem-L-free_wave_eq}]
\begin{proof}
  We start again, as in proof of lemma \ref{Lem-free_wave_eq}, with the formula
\begin{equation} \label{Duhamel-L}
  v(t,x)=\frac{1}{4\pi t}\int_{S(x,t)} {g(y)}\, d\sigma(y)
  +\frac{1}{4\pi t^2}\int_{S(x,t)} [(y-x)\cdot\nabla f(y)+f(y)]\, d\sigma(y)
\end{equation}
for $t>0, x\in\R^3$, but now we cannot use the pointwise estimates on $f$ and $g$. We must show that the integrals over the spheres make sense almost everywhere and that the resulting $v$ belongs to $\Linftx{m-1}$. 

For the moment let's forget the function $f$ and concentrate on the first integral only. 

We can cover $\R^3$ with spheres $S(x,t)$ of same origin $x$ and various radii $t\in\R_+$, i.e. $\R^3 \cong S(x,\cdot)\times \R_+$ up to one point $x\in\R^3$, which itself plays no role in integration and measurability. Since $g\in\Linfx{m}$ and $\|g\|_\Linfx{m} = \|\<\cdot\>^m g\|_\Linf = g_0 <\infty$ we have $|g(x)|\leq g_0/\<x\>^m$ a.e. on $\R^3$. The set
\begin{equation}
  A:=\{x: |g(x)|>g_0/\<x\>^m\}
\end{equation}
has therefore measure zero in $\R^3$. From a variation of Fubini's theorem\footnote{By ``variation of Fubini's theorem" we mean here the non-standard decomposition of $\R^3\cong S(x,\cdot)\times \R_+$. It is equivalent to introduction of $\widetilde{g}_x(\omega,t):=g(x+\omega t)$ defined on $S(0,1)\times \R_+$, where the standard Fubini's theorem holds.} for sets it follows that for almost all $t\in\R_+$ the measure (on the spheres $S(x,t)$) of $A|_{S(x,t)}$ is zero. In other words, for almost all $t\in\R_+$ and all $x\in\R^3$ it holds
\begin{equation} \label{est-g}
  |g(y)|\leq g_0/\<y\>^m
\end{equation}
for almost all $y\in S(x,t)$, what can be written as 
$\|\<\cdot\>^m g\|_{L^\infty_t(\R_+)\Linf(S(x,t))}\leq g_0$.


Since $g\in\Linfx{m}$ then also $g\in\Linf(\R^3)$ and $g\in L^1_\text{loc}(\R^3)$. From $\R^3\cong S(x,\cdot)\times \R_+$ and a variation of Fubini's theorem\footnotemark[\value{footnote}] it follows that for almost all $t\in\R_+$ and all $x\in\R^3$ we have $g|_{S(x,t)}\in L^1(S(x,t))$ (we skip "local" because $S(x,t)$ is compact in $\R^3$), i.e. $\int_{S(x,t)} |g(y)|\,d\sigma(y)$ exists and is finite. 
Therefore, for these $(t,x)$, using \refa{est-g}, we can estimate
\begin{equation}
  \left| \int_{S(x,t)} g(y)\,d\sigma(y)\right| \leq
  \int_{S(x,t)} |g(y)|\,d\sigma(y) \leq
  g_0 \int_{S(x,t)} \frac{d\sigma(y)}{\<y\>^m}
\end{equation}
Using the estimate \refa{est-int1} proved in lemma \ref{Lem-int}, with $m>3$, we obtain a bound for the first term in \refa{Duhamel-L}
\begin{equation}
  \left|\frac{1}{4\pi t} \int_{S(x,t)} g(y)\,d\sigma(y)\right| \leq
  g_0 \frac{C^{(1)}_m}{\norm{t+|x|}\norm{t-|x|}^{m-2}} 
\end{equation}
for almost all $t\in\R_+$ and all $x\in\R^3$.

We proceed analogously with the second term in \refa{Duhamel-L}, with $m>3$, using the estimate \refa{est-int2}  from lemma \ref{Lem-int} for the term containing $f$, and obtain
\begin{equation}
  \left|\frac{1}{4\pi t^2} \int_{S(x,t)} (y-x)\cdot \nabla f(y)\,d\sigma(y)\right| \leq
  \frac{1}{4\pi t} \int_{S(x,t)} |\nabla f(y)|\,d\sigma(y) \leq
  f_1 \frac{C^{(1)}_m}{\norm{t+|x|}\norm{t-|x|}^{m-2}} 
\end{equation}
\begin{equation}
  \left|\frac{1}{4\pi t^2} \int_{S(x,t)} f(y)\,d\sigma(y)\right| \leq
  f_0 \frac{C^{(2)}_m}{\norm{t+|x|}\norm{t-|x|}^{m-2}} 
\end{equation}
for almost all $t\in\R_+$ and all $x\in\R^3$. The last three estimates lead to
\begin{equation}
  |v(t,x)| \leq \frac{C_m\cdot(g_0 + f_1 + f_0)}{\norm{t+|x|}\norm{t-|x|}^{m-2}}
\end{equation}
with $C_m$ defined in \refa{Cp}, for almost all $t\in\R_+$ and all $x\in\R^3$. The inclusion of the boundary $t=0$ is trivial, because $v(0,x)=0$, although not necessary, since the set $\{0\}\times\R^3$ is of measure zero in $\RTXb$. Thus
\begin{equation} \label{v-Linf}
  \|v(t,x)\|_\Linftx{m-1} = 
  \|{\norm{t+|x|}\norm{t-|x|}^{m-2}} v(t,x)\|_\LinfTXb \leq C_m.
\end{equation}

It remains to show that $v$ is measurable on $\RTXb$. A product of two measurable functions is measurable, hence $g(y)$ as well as $(y-x)\nabla f$ and $f$ are measurable on $\R^3$. The solution $v(t,x)$, given by \refa{Duhamel-L}, can be rewritten as
\begin{equation}
  v(t,x)=\int_{S(0,1)} \frac{t g(x+t\omega) + 
  t\omega\cdot\nabla f(x+t\omega)+f(x+t\omega)}{4\pi}\, d\sigma(\omega)
\end{equation}
where we have changed variables so that we integrate over a unit sphere around the origin. Define $h(t,x,\omega):=[t g(x+t\omega)+t\omega\cdot\nabla f(x+t\omega)+f(x+t\omega)]/4\pi$, which is measurable on $\Rb_+\times\R^3\times S(0,1)$, because it involves only sums, products and compositions of measurable functions. Then,
\begin{equation}
  v(t,x)= \int_{S(0,1)} h(t,x,\omega)\, d\sigma(\omega),
\end{equation}
according to the theorem of Tonelli, is a measurable function of $(t,x)\in \Rb_+\times\R^3$ defined almost everywhere. Together with boundedness almost everywhere, shown in \refa{v-Linf}, we conclude that $v\in \Linftx{m-1}$.

Since $v$ is defined by \refa{Duhamel-L} almost everywhere and $v\in\Lloc$, by lemma \ref{Lem-weak-sol}, $v$ solves weakly the free wave equation \refa{lem-wave-eq-L}.
\end{proof}

\Proof[Lemma \ref{Lem-Georgiev}]
\begin{proof}
$v(t,x)$, given for $t>0, x\in\R^3$ by the formula
\begin{equation}
\begin{split}
  v(t,x) &:= \frac{1}{4\pi}\int_{K(x,t)} \frac{F(s,y)}{t-s}\;d\kappa(s,y)
\end{split}
\end{equation}
for $F\in\C^0(\RTXb)$, is well-defined everywhere in $(t,x)\in\Rb_+\times\R^3=\RTXb$, because the integral over a compact set $K(x,t)\in\RTXb$ of a continuous integrand exists and is finite.
Then, using a variation of Fubini's theorem, we split the integration
\begin{equation}
\begin{split}
  v(t,x) &:= \frac{1}{4\pi} \int_0^t ds \int_{S(x,t-s)} d\sigma(y) \frac{F(s,y)}{t-s} \\
  &= \frac{1}{4\pi} \int_0^t dt \int_{S(0,1)} d\sigma(\omega)\; (t-s) F(s,x+(t-s)\omega).
\end{split}
\end{equation}
Now, $H(t,x,s,\omega):=(t-s) F(s,x+(t-s)\omega)/(4\pi)$ is continuous in all variables $t,x,s,\omega$ and for all $(s,\omega)\in[0,T]\times S(0,1)$ is uniformly bounded on $(t,x)\in [0,T]\times \R^3$ by $|H(t,x,s,\omega)|\leq T F_0 \in L^1([0,T]\times S(0,1))$ for any $T>0$. Then, by a standard theorem on Lebesgue integration of continuous functions, the integral
\begin{equation}
  v(t,x)= \int_0^t ds \int_{S(0,1)} d\sigma(\omega)\; H(t,x,s,\omega)
\end{equation}
is continuous on $(t,x)\in [0,T]\times \R^3$ for any $T>0$ and hence on $(t,x)\in\RTXb$. (The case $t=0$ is trivially included, because $\lim_{t\ra 0^+} v(t,x)=0=v(0,x)$.) 

Since $v$ is defined everywhere in $\RTX$ and hence almost everywhere in $\RTXb$ by the above formula, and $v\in\Lloc$, then by lemma \ref{Lem-weak-sol} it solves weakly the  wave equation \refa{wave-eq-F}.

Now, knowing that $v(t,x)$ is continuous and hence finite everywhere, we can estimate
\begin{equation}
\begin{split}
  |v(t,x)| &\leq  \frac{1}{4\pi}\int_{K(x,t)} \frac{|F(s,y)|}{|t-s|} d\kappa(s,y) \\
  &\leq \frac{F_0}{4\pi} 
  \int_{K(x,t)} \frac{d\kappa(s,y)}{\norm{y}^q\norm{s+|y|}\norm{s-|y|}^{p-1} (t-s)}
\end{split}
\end{equation}
Lemma \ref{Lem-int-cone} gives an estimate for this double integral (integral over a cone) and we obtain a bound
\begin{equation}
  |v(t,x)| \leq \frac{C_{p,q}\cdot F_0}{\norm{t+|x|} \norm{t-|x|}^{p-1}}
\end{equation}
which is valid for all $(t,x)\in\RTX$. This bound is also satisfied for $t=0$, because $v(0,x)=0$.
\end{proof}

\Proof[Lemma \ref{Lem-L-Georgiev}]
\begin{proof}
We have $\norm{x}^q F\in\Linftx{p}$ then also $F\in\Linf(\RTXb)$ and hence $F\in L^1_\text{loc}(\RTXb)$. The whole space-time $\RTXb$ can be covered by the set of cones $\{K(x,t): t\in\R_+\}$, up to one point $(0,x)\in\R_+\times\R^3$, which itself plays no role in integration and measurability, i.e. $\RTXb\cong C(x,\cdot)\times \R_+$. Then, from a variation of Fubini's theorem\footnote{By ``variation of Fubini's theorem" we mean here the non-standard decomposition of $\RTX\cong C(x,\cdot)\times \R_+$. It is equivalent to introduction of $\widetilde{F}_{t,x}(\omega,s):=F(t-s,x+\omega s))$ defined on $S(0,1)\times [0,t]$, where the standard Fubini's theorem holds.} it follows that $F|_{K(x,t)}\in L^1(K(x,t))$ (we skip "local" because $K(x,t)$ is compact in $\RTXb$) for almost all $t\in\R_+$ and all $x\in\R^3$. Then, of course, $F(s,y)$ and hence $F(s,y)/(t-s)$ is measurable on the same cones, so the following integral over the cone
\begin{equation} \label{cone-int}
  v(t,x) := \frac{1}{4\pi}\int_{K(x,t)} \frac{F(s,y)}{t-s}\;d\kappa(s,y) 
\end{equation}
exists for almost all $t\in\R_+$ and all $x\in\R^3$. 

As next, its value can be estimated. The condition $\|\norm{x}^q F(t,x)\|_\Linftx{p}=F_0<\infty$ means that $|F(t,x)|\leq F_0/(\norm{x}^q\norm{t+|x|}\norm{t-|x|}^{p-1})$ for all $(t,x)\in\RTXb$ except a set $A$ of measure zero in $\RTXb$. From a variation of Fubini's theorem\footnotemark[\value{footnote}] for measurable sets and the decomposition $\RTXb\cong K(x,\cdot)\times \R_+$ it follows that for almost all $t\in\R_+$ the measure (on the cone $K(x,t)$) of $A|_{K(x,t)}$ is zero. It means that for almost all $t\in\R_+$ and all $x\in\R^3$ it holds $|F(s,y)|\leq F_0/(\norm{y}^q\norm{s+|y|}\norm{s-|y|}^{p-1})$ almost everywhere on the cone $(s,y)\in K(x,t)$. For these $(t,x)$ we can estimate
\begin{equation}
\begin{split}
  |v(t,x)| &\leq  \frac{1}{4\pi}\int_{K(x,t)} \frac{|F(s,y)|}{|t-s|} d\kappa(s,y)\\
  &\leq \frac{F_0}{4\pi}
  \int_{K(x,t)} \frac{d\kappa(s,y)}{\norm{y}^q\norm{s+|y|}\norm{s-|y|}^{p-1} (t-s)}
\end{split}
\end{equation}
Lemma \ref{Lem-int-cone} gives an estimate for this integral over the cone and we obtain a bound
\begin{equation}
  |v(t,x)| \leq \frac{C_{p,q}\cdot F_0}{\norm{t+|x|} \norm{t-|x|}^{p-1}}
\end{equation}
which is valid for almost all $t\in\R_+$ and all $x\in\R^3$. (This bound is also trivially valid for $t=0$, because $v(0,x)=0$.) Finally, it gives
\begin{equation}
  \|v\|_\Linftx{p} \leq C_{p,q} F_0. 
\end{equation}
In order to have $v\in\Linftx{p}$ it remains to show that $v(t,x)$ is measurable on $\RTXb$. To this end we use again a variation of Fubini's theorem with $K(x,t)\cong S(x,t-\cdot)\times \Rb_+$ for \refa{cone-int} and those $(t,x)$ where this integral exists and is finite, and write
\begin{equation}
  v(t,x) =  \frac{1}{4\pi}\int_0^t dt \int_{S(0,1)} d\sigma(\omega)\; (t-s) F(s,x+(t-s)\omega)
\end{equation}
Since $\norm{x}^q F(t,x)\in\Linftx{p}$ the function $F(t,x)$ is measurable on $\Rb_+\times\R^3$. Then $H(t,x,s,\omega):=(t-s) F(s,x+(t-s)\omega)/(4\pi)$ is measurable on $(t,x,s,\omega)\in \Rb_+\times\R^3\times\Rb_+\times S(0,1)$. From Tonelli's theorem we get
\begin{equation}
  v(t,x)= \int_0^t dt \int_{S(0,1)} d\sigma(\omega)\; H(t,x,s,\omega)
\end{equation}
measurable on $\Rb_+\times\R^3=\RTXb$. Finally, $v\in\Lloc$ and is defined almost everywhere by \refa{cone-int}, hence by Lemma \ref{Lem-weak-sol} $v$ solves weakly the wave equation \refa{wave-eq-F-L}. 
\end{proof}

\Proof[Theorem \ref{Th-Ldecay}]
\begin{proof}
Consider the following iteration scheme
\begin{equation}
  u_{n+1}:=I_0(f,g)-L_0(Vu_n), \quad n=0,1,2,...\quad \text{and}\quad u_{-1}:=0.
\end{equation}
For $g,\nabla f\in\Linfx{m}$ and $f\in\Linfx{m-1}$ with $m>3$ from lemma \ref{Lem-L-free_wave_eq} we get $u_0=I_0(f,g)\in\Linftx{m-1}$. Next, observe that if $u_n\in\Linftx{p}$ with some $p>1$ then
\begin{equation}
  \|\<x\>^k V u_n\|_\Linftx{p} \leq \|\<x\>^k V\|_\Linf \|u_n\|_\Linftx{p} = V_0 \|u_n\|_\Linftx{p}<\infty
\end{equation}
and from lemma \ref{Lem-L-Georgiev} with $F\equiv V u_n$ we get $L_0(V u_n)\in\Linftx{p}$ when $p\leq k$. Because $\Linftx{p_1}\subset\Linftx{p_2}$ when $p_1\geq p_2$, we get $u_{n+1}\in\Linftx{p}$ with $p\leq \min(m-1,k)$. By induction we obtain $u_n\in\Linftx{p}$ for every $n=0,1,2,...$ with the optimal value $p:=\min(m-1,k)$. Then, we have
\begin{equation}
\begin{split}
  \|u_{n+1}-u_n\|_\Linftx{p} &= \|L_0(-V(u_n-u_{n-1}))\|_\Linftx{p} \leq
  C_{p,k} \|\<x\>^k V (u_n-u_{n-1})\|_\Linftx{p} \\ &\leq 
  C_{p,k} \|\<x\>^k V\|_\Linf \|u_n-u_{n-1}\|_\Linftx{p} =
  C_{p,k} V_0 \|u_n-u_{n-1}\|_\Linftx{p}
\end{split}
\end{equation}
again making use of lemma \ref{Lem-L-Georgiev} with $F\equiv -V (u_n-u_{n-1}) \in \<x\>^{-k}\Linftx{p}$.
For $\delta:=C_{p,k} V_0<1$ the iteration is a contraction in Banach space $\Linftx{p}$. A simple argument shows that the sequence $u_n$ is Cauchy. We have 
\begin{equation}
  \|u_{k+1}-u_k\|_\Linftx{p} \leq \delta^{k+1} \|u_0-u_{-1}\|_\Linftx{p} = 
  \delta^{k+1} \|I_0(f,g)\|_\Linftx{p}
\end{equation}
and for $n>m$
\begin{equation}
\begin{split}
  \|u_n-u_m\|_\Linftx{p} &\leq \sum_{k=0}^{n-m-1} \|u_{m+k+1}-u_{m+k}\|_\Linftx{p} 
  \leq \sum_{k=0}^{n-m-1} \delta^{k+m+1} \|I_0(f,g)\|_\Linftx{p}\\
  &\leq \frac{\delta^{m+1}}{1-\delta} \|I_0(f,g)\|_\Linftx{p}
\end{split}
\end{equation}
and this expression can be made arbitrarily small ($<\epsilon$) for all $n,m>M(\epsilon)$ when $\delta<1$. Since $\Linftx{p}$ is a Banach space the Cauchy sequence $u_n$ has a limit $u\in\Linftx{p}$ with the property
\begin{equation} \label{u-sol}
  u=I_0(f,g)-L_0(Vu)
\end{equation}
and hence 
\begin{equation}
  \|u\|_\Linftx{p}\leq \|I_0(f,g)\|_\Linftx{p}+\|L_0(Vu)\|_\Linftx{p}
  \leq C_m(f_0+f_1+g_0)+C_{p,k} V_0\|u\|_\Linftx{p}.
\end{equation}
Then
\begin{equation}
  \|u\|_\Linftx{p}\leq \frac{C_m(f_0+f_1+g_0)}{1-C_{p,k} V_0} \equiv C.
\end{equation}
Moreover, equation \refa{u-sol} together with lemmas \ref{Lem-free_wave_eq} and \ref{Lem-Georgiev} imply
\begin{equation}
\begin{split}
  \int dt \int d^3x\; \Box \varphi(t,x) \; u(t,x) =& 
  - \int d^3x\; \d_t \varphi(0,x)\; f(x) + \int d^3x\; \varphi(0,x)\; g(x)\\
  &- \int_{\R} dt \int d^3x\;  \varphi(t,x)\; V(x)\; u(t,x),
 \end{split}
\end{equation}
what gives equation \refa{weak-sol-pot}. Since $u\in\Lloc$ it is a weak solution to the wave equation \refa{wave-eq}. Uniqueness is guaranteed by theorem \ref{Th-Energy}.
\end{proof}

\Proof[Theorem \ref{Th-decay}]
\begin{proof}
The condition $|u(t,x)|\leq C/(\norm{t+|x|}\norm{t+|x|}^{p-1})$ implies $u\in\Linftx{p}$ and together with $u\in\C^0(\RTXb)$ these two conditions are equivalent. Therefore, proof of this theorem is analogous to the proof of theorem \ref{Th-Ldecay} with the only difference that continuity must be shown separately in the iteration. 

Analogously, for iteration defined by
\begin{equation}
  u_{n+1}:=I_0(f,g)-L_0(Vu_n), \quad n=0,1,2,...\quad \text{and}\quad u_{-1}:=0.
\end{equation}
and $(f,g)\in\C^1(\R^3)\times\C^0(\R^3)$ satisfying the bounds \refa{f-g-bound}, hence $g,\nabla f\in\Linfx{m}$, by lemma \ref{Lem-free_wave_eq}, we get $u_0=I_0(f,g)\in\Linftx{m-1}\cap \C^0(\RTXb)$. Next, if $u_n\in\Linftx{p}\cap\C^0(\RTXb)$ for some $p>2$, hence also $Vu_n\in\C^0(\RTXb)$, because $V\in\C^0(\R^3)$, then analogously, by lemma \ref{Lem-Georgiev} with $F\equiv V u_n$, we get $L_0(V u_n)\in\Linftx{p}\cap \C^0(\RTXb)$ when $p\leq k$. It implies that $u_{n+1}\in\Linftx{p}\cap \C^0(\RTXb)$ with $p\leq \min(m-1,k)$. By induction we obtain $u_n\in\Linftx{p}\cap \C^0(\RTXb)$ for every $n=0,1,2,...$ with the optimal value $p:=\min(m-1,k)$. Analogously, using lemma \ref{Lem-Georgiev} with $F\equiv V (u_n-u_{n-1}) \in \<x\>^{-k}\Linftx{p}\cap \C^0(\RTXb)$ we show
\begin{equation}
  \|u_{n+1}-u_n\|_\Linftx{p} \leq C_{p,k} V_0 \|u_n-u_{n-1}\|_\Linftx{p}
\end{equation}
what presents a contraction in Banach space $\Linftx{p}\cap \C^0(\RTXb)$. Again by full analogy, for $V_0$ small enough, we show that the sequence $u_n$ is Cauchy in Banach space $\Linftx{p}\cap \C^0(\RTXb)$, hence $u_n$ has a limit $u\in\Linftx{p}\cap \C^0(\RTXb)$ with the property
\begin{equation} \label{u-sol-L}
  u=I_0(f,g)-L_0(Vu).
\end{equation}
Then
\begin{equation}
  \|u\|_\Linftx{p}\leq \frac{C_m(f_0+f_1+g_0)}{1-C_{p,k} V_0} \equiv C.
\end{equation}
Since $u\in\C^0(\RTXb)$ the last condition implies
\begin{equation}
  |u(t,x)|\leq \frac{C}{\norm{t+|x|}\norm{t+|x|}^{p-1}}\qquad\forall (t,x)\in\RTXb.
\end{equation}
And also analogously, equation \refa{u-sol-L} together with lemmas \ref{Lem-L-free_wave_eq} and \ref{Lem-L-Georgiev} imply
\begin{equation}
\begin{split}
  \int dt \int d^3x\; \Box \varphi(t,x) \; u(t,x) =& 
  - \int d^3x\; \d_t \varphi(0,x)\; f(x) + \int d^3x\; \varphi(0,x)\; g(x)\\
  &- \int_{\R} dt \int d^3x\;  \varphi(t,x)\; V(x)\; u(t,x),
 \end{split}
\end{equation}
what gives equation \refa{weak-sol-pot}. Since $u\in\Lloc$ it is a weak solution to the wave equation \refa{wave-eq}. Uniqueness is guaranteed by theorem \ref{Th-Energy}.
\end{proof}

\Proof[Corollary \ref{Cor-wave-eq-V-F}]
\begin{proof}
  Analogously as in proofs of the theorems above we define an iteration 
\begin{equation}
  u_{n+1}:=I_0(f,g)-L_0(Vu_n)-L_0(F), \quad n=0,1,2,...\quad \text{and}\quad u_{-1}:=0.
\end{equation}
From lemma \ref{Lem-L-free_wave_eq} we get $u_0=I_0(f,g)\in\Linftx{m-1}$. 
From lemma \ref{Lem-L-Georgiev} we get $L_0(F)\in\Linftx{r}$.
By same argument as in previous theorems we have $L_0(V u_n)\in\Linftx{p}$ with $p\leq k$ provided $u_n\in\Linftx{p}$ with some $p>1$. 
Then, by induction we obtain $u_n\in\Linftx{p}$ for every $n=0,1,2,...$ with the optimal value $p:=\min(m-1,k,r)$. Then, we find
\begin{equation}
\begin{split}
  \|u_{n+1}-u_n\|_\Linftx{p} &= \|L_0(-V(u_n-u_{n-1}))\|_\Linftx{p} \leq
  C_{p,k} \|\<x\>^k V (u_n-u_{n-1})\|_\Linftx{p} \\ &\leq 
  C_{p,k} \|\<x\>^k V\|_\Linf \|u_n-u_{n-1}\|_\Linftx{p} =
  C_{p,k} V_0 \|u_n-u_{n-1}\|_\Linftx{p}
\end{split}
\end{equation}
again making use of lemma \ref{Lem-L-Georgiev} with $F\equiv -V (u_n-u_{n-1}) \in \<x\>^{-k}\Linftx{p}$.
For $\delta:=C_{p,k} V_0<1$ the iteration is a contraction in Banach space $\Linftx{p}$ and the sequence $u_n$ is Cauchy. It converges to the limit $u\in\Linftx{p}$ (Banach space) which satisfies
\begin{equation} 
  u=I_0(f,g)-L_0(Vu)-L_0(F)
\end{equation}
and hence 
\begin{equation}
\begin{split}
  \|u\|_\Linftx{p} &\leq \|I_0(f,g)\|_\Linftx{p}+\|L_0(Vu)\|_\Linftx{p}+\|L_0(F)\|_\Linftx{p}\\
  &\leq C_p(f_0+f_1+g_0)+C_{p,k} V_0\|u\|_\Linftx{p} + C_{r,q} F_0.
\end{split}
\end{equation}
Then
\begin{equation}
  \|u\|_\Linftx{p}\leq \frac{C_m(f_0+f_1+g_0)+C_{r,q} F_0}{1-C_{p,k} V_0} \equiv C.
\end{equation}

To prove the continuous case one needs only to use the lemmas \ref{Lem-free_wave_eq} and \ref{Lem-Georgiev} instead of the lemmas \ref{Lem-L-free_wave_eq} and \ref{Lem-L-Georgiev} and repeat the convergence argument from the proof of theorem \ref{Th-decay}.
\end{proof}

\section{Decay estimates for the derivatives} \label{Sec-Derivatives}

One can state the estimate given in theorem \ref{Th-decay} or \ref{Th-Ldecay} in a more detailed form, namely that the solution of \refa{wave-eq} satisfies
\begin{equation}
  \|u\|_\Linftx{p} \leq \frac{C_m\cdot(\|f\|_\Linfx{m-1} + \|\nabla f\|_\Linfx{m} 
  + \|g\|_\Linfx{m})}{1-C_{p,k}\cdot \|V\|_\Linfx{k}}
\end{equation}
when $k>2$, $m>3$, $p:=\min(k,m-1)$, the norms $\|f\|_\Linfx{m-1}, \|\nabla f\|_\Linfx{m}, \|g\|_\Linfx{m}$ are finite and $\|V\|_\Linfx{k} < C_{p,k}^{-1}<\infty$.

If $u$ is a classical solution one can differentiate the wave equation \refa{wave-eq} with respect to time and obtain
\begin{equation} 
  \Box \d_t{u} + V \d_t{u} = 0.
\end{equation}
If $u$ is a weak solution of \refa{wave-eq} then this equation is also satisfied in the weak sense. The existence of the weak solution will be guaranteed by the theorems below.
It can be treated again as equation \refa{wave-eq} for a new variable $v:=\d_t u$ with the initial data
\begin{equation}
  v(0,x)=g(x),\qquad \d_t v(0,x)=\d^2_t u(0,x) = -A f,
\end{equation}
where $A:=-\lap+V(x)$. Then, theorem \ref{Th-decay} or \ref{Th-Ldecay} gives existence of the solution $\d_t u\in \C^0(\RTXb)$ or $\d_t u\in \Linftx{p'}$, respectively, and the estimate
\begin{equation}
  \|\d_t u\|_\Linftx{p'} \leq \frac{C_m\cdot(\|g\|_\Linfx{m'-1} + \|\nabla g\|_\Linfx{m'} 
  + \|A f\|_\Linfx{m'})}{1-C_{p',k'}\cdot \|V\|_\Linfx{k}}.
\end{equation}
The term $\|A f\|_\Linfx{m'}$ can be bounded by $\|\lap f\|_\Linfx{m'} + \|V f\|_\Linfx{m'}$. Analogously, one can obtain existence and an estimate for $\d_t^2 u$
\begin{equation}
  \|\d_t^2 u\|_\Linftx{p''} \leq \frac{C_m\cdot(\|A f\|_\Linfx{m''-1} + \|\nabla (A f)\|_\Linfx{m''} 
  + \|A g\|_\Linfx{m''})}{1-C_{p'',k''}\cdot \|V\|_\Linfx{k}}.
\end{equation}
The term $\|\nabla (A f)\|_\Linfx{m''}$ can be bounded by $\|\nabla \lap f\|_\Linfx{m''}+\|V \nabla f\|_\Linfx{m''}+\|(\nabla V) f\|_\Linfx{m''}$. The three estimates can be put together when the constants $m,m',m''$ and $p,p',p''$ are related to each other. Using the following property of the weighted $\Linf$ norms $\|h_1\cdot h_2\|_\Linfx{a+b} \leq \|h_1\|_\Linfx{a}\cdot\|h_2\|_\Linfx{b}$ 
we get for $m=m'=m''>3$
\renewcommand{\theCorA}{2a}
\begin{CorA} \label{Cor-derivs-time} 
\begin{equation}
  u, \d_t u, \d_t^2 u \in \Linftx{p}
\end{equation}
with $p:=\min(k,m-1)$ provided
\begin{equation}
  f, \nabla f, \lap f, \nabla \lap f, g, \nabla g, \lap g \in \Linfx{m},
\end{equation}
\begin{equation}
  \|V\|_\Linfx{k} < C_{p,k}^{-1}< \infty,\qquad 
  \|\nabla V\|_\Linf < \infty.
\end{equation}
\end{CorA}
\noindent Alternatively, choosing $m=m'-1=m''-2>3$ we get for classical solutions especially simple form of the assumptions
\renewcommand{\theCorA}{2a'}
\begin{CorA}
\begin{equation}
  \|u\|_\Linftx{p}, \|\d_t u\|_\Linftx{p'}, \|\d_t^2 u\|_\Linftx{p''} < \infty, 
\end{equation}
with $p:=\min(k,m-1)$, $p':=\min(k,m)$, $p'':=\min(k,m+1)$ provided $(f,g)\in\C^3(\R^3)\times\C^2(\R^2)$ with
\begin{equation}
  |\nabla^3 f| \in \Linfx{m+2}, \qquad |\nabla^2 g| \in \Linfx{m+2}
\end{equation}
and $V\in\C^1(\R^3)$ with
\begin{equation}
  \|V\|_\Linfx{k} < \min(C_{p,k}^{-1},C_{p',k}^{-1},C_{p'',k}^{-1}) < \infty,\qquad 
  \|\nabla V\|_\Linfx{3} < \infty.
\end{equation}
\end{CorA}
\noindent Here, we introduced a simplified notation\footnote{This notation simplifies the assumptions, but looses information about the decay of directional derivatives, which is however rarely used.} 
\begin{equation}
  |\nabla^n h|:=\sum_{a_1,...,a_n=1}^3 |\d_{a_1} ... \d_{a_n} h|.
\end{equation}
Because of the regularity, the bounds on $|\nabla^3 f|$ and $|\nabla^2 g|$ can be integrated to give
\begin{equation}
  \|f\|_\Linfx{m-1}, \|\nabla f\|_\Linfx{m}, \|\lap f\|_\Linfx{m+1}, \|\nabla \lap f\|_\Linfx{m+2} < \infty,
\end{equation}
and
\begin{equation}
  \|g\|_\Linfx{m}, \|\nabla g\|_\Linfx{m+1}, \|\lap g\|_\Linfx{m+2} < \infty.
\end{equation}

The control of spatial derivatives is more difficult, since $V$ depends on $x$. Again, by differentiation of the wave equation  \refa{wave-eq} with respect to $x^i$ we get (for weak solutions the equation holds as well by the same argument as above)
\begin{equation} 
  \Box \d_i{u} + V \d_i{u} = - (\d_i V) u.
\end{equation}
This equation can be handled with Corollary \ref{Cor-wave-eq-V-F} for $v=\d_i u$ and $F=- (\d_i V) u$ with the initial data $v(0,x)=\d_i f(x)$ and $\d_t v(0,x) = \d_i g(x)$. Then, 
\begin{equation}
  \|\d_i u\|_\Linftx{p'}\leq 
  \frac{C_{m'}(\|\d_i f\|_\Linfx{m'-1}+\|\nabla \d_i f\|_\Linfx{m'}+\|\d_i g\|_\Linfx{m'})
  +C_{p,k'} \|\norm{x}^{k'} (\d_i V) u\|_\Linftx{p}}{1-C_{p',k} \|V\|_\Linfx{k}}.
\end{equation}
for $p':=\min(k,m'-1,p)$ and $m'>3, k,k'>2, k'\geq p>1$.
The last term in the nominator can be bounded by $\|(\d_i V)\|_\Linfx{k'} \| u\|_\Linftx{p}$.
Analogously, an estimate for $\d_i\d_k u$ reads
\begin{multline}
  \|\d_i\d_k u\|_\Linftx{p''}\leq 
  \Big[C_{m''}(\|\d_i\d_k f\|_\Linfx{m''-1}+\|\nabla \d_i\d_k f\|_\Linfx{m''}
  +\|\d_i\d_k g\|_\Linfx{m''}) \\ 
  +C_{p,k''} \|(\d_i\d_k V)\|_\Linfx{k''} \| u\|_\Linftx{p}
  +2 C_{p',k'} \|\d V\|_\Linfx{k'} \|\d u\|_\Linftx{p'}\Big]/
  \big[1-C_{p'',k} \|V\|_\Linfx{k}\big]
\end{multline}
for $p'':=\min(k,m''-1,p,p')$ and $m''>3, k,k',k''>2, k''\geq p>1, k'\geq p'>1$. We can combine these estimates for $m=m'=m''>3$ and $k=k'=k''>2$ and get
\renewcommand{\theCorA}{2b}
\begin{CorA} \label{Cor-derivs-spatial} 
\begin{equation}
  u, |\nabla u|, |\nabla^2 u| \in \Linftx{p}  
\end{equation}
with $p:=\min(k,m-1)$ provided
\begin{equation}
  f\in \Linfx{m-1},\qquad  
  |\nabla f|, |\nabla^2 f|, |\nabla^3 f|, g, |\nabla g|, |\nabla^2 g| \in \Linfx{m},
\end{equation}
\begin{equation}
  V, |\nabla V|, |\nabla ^2 V| \in \Linfx{k} \quad \text{and} \quad
  \|V\|_\Linfx{k} < C_{p,k}^{-1}< \infty.
\end{equation}
\end{CorA}
Finally, combining corollary \ref{Cor-derivs-time} with \ref{Cor-derivs-spatial} we get corollary \ref{Cor-decay-derivs}.

The same reasoning can be repeated for $(f,g)\in\C^3(\R^3)\cap\C^2(\R^3)$ and $V\in\C^2(\R^3)$ which give the classical solution $u\in\C^2(\RTXb)$. The only difference is that one needs to use theorem \ref{Th-decay} instead of \ref{Th-Ldecay} and the continuous version of corollary \ref{Cor-wave-eq-V-F}.

Higher derivatives can be treated in an analogous way.







\bibliography{QNMs}

\begin{thebibliography}{10}

\bibitem{Strauss-T}
W.~Strauss and K.~Tsutaya.
\newblock Existence and blow up of small amplitude nonlinear waves with a
  negative potential.
\newblock {\em Discr. Cont. Dynamical Systems}, 3(2):175--188, 1997.

\bibitem{ChingComplPRL}
E.~S.~C. Ching, P.~T. Leung, W.~M. Suen, and Young K.
\newblock Quasinormal mode expansion for linearized waves in gravitational
  systems.
\newblock {\em Phys. Rev. Lett.}, 74(23):4588--4591, 1995.

\bibitem{ChingTails}
E.~S.~C. Ching, P.~T. Leung, W.~M. Suen, and Young K.
\newblock Wave propagation in gravitational systems: Late time behavior.
\newblock {\em Phys. Rev. D}, 52(4):2118--2132, 1995.

\bibitem{Hod-tails}
S.~Hod.
\newblock Wave tails in non-trivial backgrounds.
\newblock {\em Class. Quantum Grav.}, 18:1311--1318, 2001.
\newblock gr-qc/0008001.

\bibitem{Georg-H-K}
V.~Georgiev, Ch. Heiming, and H.~Kubo.
\newblock Supercritical semilinear wave equation with non-negative potential.
\newblock {\em Comm. Partial Diff. Eq.}, 11-12(26):2267--2303, 2001.

\bibitem{RS-IV}
M.~Reed and B.~Simon.
\newblock {\em Methods of Modern Mathematical Physics}, volume IV: Analysis of
  Operators.
\newblock Academic Press, New York, San Francisco, London, 1978.

\bibitem{Thoe-WavPot}
D.~Thoe.
\newblock Spectral theory for the wave equation with a potential term.
\newblock {\em Arch. Rational Mech. Anal.}, (22):364--406, 1966.

\bibitem{Asakura}
Asakura F.
\newblock Existence of a global solution to a semi-linear wave equation with
  slowly decreasing initial data in three space dimenstions.
\newblock {\em Comm. Part. Diff. Eq.}, 13(11):1459--1487, 1986.

\bibitem{John-blowup}
F.~John.
\newblock Blow-up of solutions of nonlinear wave equations in three space
  dimensions.
\newblock {\em Manuscripta Math.}, (28):235--268, 1979.

\bibitem{RS-II}
M.~Reed and B.~Simon.
\newblock {\em Methods of Modern Mathematical Physics}, volume II: Fourier
  Analysis, Self-Adjointness.
\newblock Academic Press, New York, San Francisco, London, 1975.

\bibitem{Wouk}
A.~Wouk.
\newblock A note on square roots of positive operators.
\newblock {\em SIAM Review}, 8(1):100--102, 1966.

\bibitem{Burq}
N.~Burq, F.~Planchon, J.G. Stalker, and A.S. Tahvildar-Zadeh.
\newblock Strichartz estimates for the wave and {S}chr{\"o}dinger equations
  with potentials of critical decay.
\newblock {\em Indiana Univ. Math. J.}, 53(6):1665--1680, 2004.
\newblock math.AP/0401019.

\bibitem{Goldstein}
J.~Goldstein.
\newblock {\em Semigroups of Linear Operators and Applications}.
\newblock Oxford University Press, 1985.
\newblock Chapter 7.

\bibitem{Taylor-I}
M.~E. Taylor.
\newblock {\em Partial Differential Equations}, volume I: Basic Theory.
\newblock Springer-Verlag New York, Inc., 1996.

\bibitem{Sogge-book}
Ch.D. Sogge.
\newblock {\em Lectures on nonlinear wave equations}.
\newblock Monographs in Analysis. Volume II. International Press Inc., 1995.

\bibitem{Rudin}
W.~Rudin.
\newblock {\em Functional Analysis}.
\newblock second edition, McGraw-Hill, Inc., 1991.
\newblock Chapter 6.

\end{thebibliography}
\bibliographystyle{unsrt}

\end{document}